# Signatures of spin-polarized p-wave superconductivity in the kagome material $RbV_3Sb_5$

Shuo Wang[1,3,4,8], Xilin Feng[2,8], Jing-Zhi Fang[1,3,8], Jia-Peng Peng[1,3], Zi-Ting Sun[2], Jia-Jie Yang[1,3], Jingchao Liu[5], Jia-Ji Zhao[6], Ning Kang[7], Xiao-Song Wu[6], Zhensheng Zhang[1,3], Xuewen Fu[5], Kam Tuen Law[2,†], Ben-Chuan Lin[1,3,4,*], Dapeng Yu[1,3,4,#]

[1] *International Quantum Academy, Shenzhen 518048, China.*

[2]*Department of Physics, Hong Kong University of Science and Technology, Clear Water Bay, Hong Kong, China.*

[3]*Shenzhen Institute for Quantum Science and Engineering, Southern University of Science and Technology, Shenzhen, 518055, China.*

[4]*Guangdong Provincial Key Laboratory of Quantum Science and Engineering, Southern University of Science and Technology, Shenzhen, 518055, China.*

[5]*Ultrafast Electron Microscopy Laboratory, The MOE Key Laboratory of Weak-Light Nonlinear Photonics, School of Physics, Nankai University, Tianjin 300071, China.*

[6]*State Key Laboratory for Artificial Microstructure and Mesoscopic Physics, Frontiers Science Center for Nano-optoelectronics, Peking University, Beijing 100871, China.*

[7]*Key Laboratory for the Physics and Chemistry of Nanodevices, School of Electronics, Peking University, Beijing 100871, China.*

[8]*These authors contributed equally to this work.*

[†]phlaw@ust.hk
[*]linbc@sustech.edu.cn
[#]yudp@sustech.edu.cn

**Abstract**

The study of kagome materials[1–4] has attracted much attention in the past few years due to the presence of many electron-electron interaction-driven phases in a single material. These include charge density waves[5–16], nematic phases[10,14], superconducting phases[11,17–25], and pair density waves[17,23,24]. In this work, we report the discovery of intrinsic spin-polarized p-wave superconductivity in the thin-flake kagome material $RbV_3Sb_5$. Firstly, when an in-plane magnetic field is swept in opposite directions, we observe a unique form of hysteresis in magnetoresistance which is different from the hysteresis induced by extrinsic mechanisms such as flux-trapping[26–28] or superheating and supercooling effects[29,30]. The unconventional hysteresis indicates the emergence of an intrinsic time-reversal symmetry-breaking superconducting phase. Strikingly, at a fixed magnetic field, the finite-resistance state can be transitioned into the superconducting state by applying and subsequently removing a large current. Secondly, at temperatures around 400 mK, the re-entrance of superconductivity occurs during an in-plane field-sweeping process with a fixed sweeping direction. This kind of re-entrance is asymmetric about the zero field axis and observed in all field directions for a fixed current direction, which is different from the re-entrance observed in conventional superconductors[31–33]. Moreover, the angle-dependent in-plane critical field measurements reveal a two-fold symmetry that deviates from the original, centrosymmetric $D_{6h}$ point group symmetry of the crystal. These findings put very strong constraints on the possible superconducting pairing symmetry of $RbV_3Sb_5$. We point out that the pairing symmetry, which is consistent with the crystal symmetry and all the observed novel properties, is possibly a time-reversal symmetry-breaking, p-wave pairing with net spin polarization[34–36]. Importantly, this p-wave pairing gives rise to a nodal topological superconducting state with Majorana flat bands on the sample edges.

The discovery of kagome superconductors[1–4] $AV_3Sb_5$ (A=K, Rb, Cs) has garnered significant attention in recent years due to their intriguing interplay of superconductivity[11,17–25], charge density waves[5–16], nematic orders[10,14], pair density waves[17,23,24], all within a single material. Such intricate interplay of the intertwined orders holds promise for uncovering unconventional phenomena, notably unconventional superconductivity[11,17–25]. Indeed, the nodal superconducting states have been observed in the kagome superconductor $RbV_3Sb_5$ with μSR experiments[20], while nodeless superconductivity was suggested in the $CsV_3Sb_5$ compound[37–40]. Up to now, the mechanism of the superconducting properties of the $AV_3Sb_5$ family remains an open question[4], necessitating further investigation and exploration.

Here, we provide evidence of time-reversal symmetry-breaking, spin polarized p-wave superconductivity in the kagome material $RbV_3Sb_5$. The observed superconducting states in $RbV_3Sb_5$ are characterized by the emergence of the in-plane magnetic field-driven hysteresis in the resistance measurements. In addition, by heating the sample with a large current and subsequently removing it, all at a fixed magnetic field within the hysteresis loop, the sample can be brought from a finite-resistance state to a superconducting state. Moreover, an asymmetric re-entrance of superconductivity driven by the in-plane magnetic field is observed at around 400 mK, showing two superconducting domes in the temperature-magnetic field phase diagram. Through magnetotransport experiments, we observe a distinct anisotropic two-fold symmetry of the in-plane upper critical fields in this centrosymmetric material. These findings put strong constraints on the potential pairing symmetry of the superconducting state. We show that all the above-mentioned novel properties of the kagome superconductor $RbV_3Sb_5$ can be understood if the pairing is a time-reversal symmetry-breaking, p-wave pairing state with a net spin polarization. The d-vector, which characterizes the pairing order parameter, has the property that $i\vec{d}_{\vec{k}} \times \vec{d}^{*}_{\vec{k}} \neq 0$, indicating a net spin-polarization in the superconducting state[34–36]. Importantly, this pairing belongs to the $B_{2u}$ or $B_{3u}$ representation of the $D_{2h}$ point group. Our theory shows that this is a nodal topological superconducting state with Majorana flat bands[41–48]. It is expected that there is a large number of zero energy Majorana modes residing on the edges of the sample. Therefore, we suggest that $RbV_3Sb_5$ is possibly a novel platform for studying the long-sought-after intrinsic topological superconductors[49–59] and zero energy Majorana flat band modes[60–64].

## Basic superconducting properties

The kagome family $AV_3Sb_5$ represents a class of layered kagome superconductors, characterized by the alternating arrangement of V-Sb slabs, Sb slabs and A slabs. Specifically, $RbV_3Sb_5$ is a member of this family, which consists of V-Sb slabs interleaved with Rb layers. Each V-Sb slab exhibits a two-dimensional (2D) kagome net of V atoms, with Sb atoms occupying central positions of hexagons within the kagome lattice, as depicted in Fig. 1a. Evidently, this structural arrangement signifies a six-fold rotational symmetry. Fig. 1b shows the calculated energy band of the kagome superconductor, highlighting an important feature near the Fermi energy: the parity mixing between the d-orbital of the V atom and the p-orbital of the Sb atom, introducing a spin-orbit-parity-coupling term[65,66] in contrast to the conventional spin-orbit-coupling term. In the two-dimensional limit, this spin-orbit-parity-coupling term is responsible for the observation of the two-fold symmetric in-plane upper critical field in this centrosymmetric material[65,66] which exhibits a nematic normal state[10]. In Fig. 1c, we present the current-voltage curves of the device at different temperatures, showcasing a characteristic superconducting behavior. Correspondingly, Fig. S1 shows the device image, in which the x-axis is roughly along the source-drain direction and the long axis of the sample. Fig. 1d depicts the typical in-plane magnetic field dependence of the differential resistance dV/dI at a temperature T = 150 mK. Apparently, the critical current decreases with the magnetic field increasing, a hallmark of superconducting behavior. The in-plane superconducting coherence length is estimated to be around 137 nm according to the linearized Ginzburg–Landau (GL) expression[67] $B_{C\perp} = \Phi_0/2\pi\xi_0^2\,(1 - T/T_C)$, where $\Phi_0$ is the magnetic flux quantum, $\xi_0$ is the zero-temperature in-plane coherence length, and $T_c$ is the critical temperature, as shown in Fig. S2.

## Hysteretic superconducting behavior

We proceed by investigating the magnetic field dependence of the kagome superconductor $RbV_3Sb_5$. The configuration of the device is depicted in Fig. 2a. We employ a three-dimensional (3D) vector magnetic field to study the magnetic response of the superconducting states. In Fig. 2a, the direction of the magnetic field is defined by the θ and $\phi$ which are the polar angle and the azimuthal angle, respectively. Specifically, $\theta = \phi = 0^{\circ}$ denotes the direction of the x-axis. Remarkably, the magnetoresistance curve exhibits pronounced hysteretic behavior, as illustrated in Fig. 2b. When an in-plane magnetic field oriented along the x-direction is swept from -1 T to 1 T (red curve), the superconducting state is initially established at around -410 mT; then as the magnetic field turns positive, the resistance deviates from zero at about 210 mT, exhibiting pronounced asymmetry. Moreover, sweeping the field from 1 T back to -1 T (black curve), produces a mirrored response, thus uncovering hysteretic behavior. If we define the critical magnetic field by the magnitude of the magnetic field at which the resistance reaches 50% of the normal state resistance, then we get the critical values at the positive magnetic field during forward (denoted as $B_{\rightarrow}$) and backward ($B_{\leftarrow}$) sweeps as $B_{\rightarrow}$ = 380 mT and $B_{\leftarrow}$ = 620 mT, respectively. Similar hysteresis is also observed when the magnetic field is applied along the y-axis as depicted in Fig. 2c. Fig. 2e and 2f further show the hysteretic behavior observed during the field sweeping from 0 T to -1 T or 1 T, respectively.

A natural question is whether such hysteretic behavior is exclusively confined to the in-plane orientation. To answer this question, we conduct the out-of-plane magnetic field angle dependence of the hysteresis. As depicted in Fig. 2g, our findings demonstrate that at an out-of-plane angle $\theta = \pm 9^{\circ}$, the hysteresis is destroyed. Such out-of-plane magnetic field angle dependence suggests that the hysteresis is related to the in-plane magnetization of the superconducting state. Consequently, this rules out the possibility that the hysteresis is triggered by the out-of-plane component of the magnetic field.

## The two-fold symmetric upper critical field

We further study the $\phi$-dependence of the magnetoresistance under the in-plane magnetic field. The magnetoresistance as a function of $\phi$ are shown in Fig. 2d. A distinct two-fold symmetric behavior is evident. The extraction of the upper critical fields is further illustrated in Fig. S3a. As shown in the theoretical calculations (Fig. S3b) within the single-layer limit, the presence of the nematic order and the spin-orbit-parity-coupling (SOPC)[65,66] can lead to a two-fold symmetric upper critical field (details can be found in the Supplementary Information) even when the orbital effects of the in-plane magnetic field are ignored. Due to the finite thickness of the sample, the orbital effects of the magnetic field can also induce this two-fold symmetry[68,69], given the six-fold rotational symmetry is broken by the nematic order in the normal state[10]. The extra contribution of the orbital effect only changes the results quantitatively. The key conclusion of observed two-fold symmetric upper critical field is that the original $D_{6h}$ point group symmetry is reduced to $D_{2h}$ due to the presence of the nematic state.

## The current heating and cooling experiment

In Fig. 2h, the result of the current heating and cooling experiment is depicted. Initially, as the magnetic field sweeps from -1 T to 270 mT (illustrated by the red curve) and the system enters into the state with non-zero resistance. At this point, we apply a large current to heat up the sample to suppress superconductivity completely. Subsequently, removing the current allows the system to cool down and enter into the superconducting state, as indicated by the red curve. Notably, with further increasing the magnetic field to 1 T, the resistance curve closely resembles the one observed during the magnetic field's reversal from 1 T to 270 mT (shown by black curve). Furthermore, by repeating the current heating and cooling experiment while sweeping the magnetic field from 1 T to -270 mT (black curve), a similar phenomenon is observed. This finding clearly shows that the finite-resistance state within the hysteresis loop is a metastable state which has a higher energy than the ground state. As will be discussed below, this observation provides crucial evidence for the existence of superconducting domains in the sample.

## Magnetic field driven re-entrance of superconductivity

Subsequently, an investigation into the temperature dependence of the magnetoresistance is studied. As temperature rises, superconductivity can be suppressed by thermal effects. Fig. 3a&b show the temperature-magnetic field phase diagram and waterfall plot of the temperature dependence of the magnetoresistance curve, respectively, while the complete temperature-magnetic field phase diagram is available in the supplementary Fig. S5a. Remarkably, within Fig.3a, two distinct superconducting domes are observed in the magnetoresistance measurements when the magnetic field changes from -1 T to 1 T. Initially, the superconductivity is destroyed at a finite magnetic field of 80 mT at 400 mK. However, a second dome of superconductivity is observed when the magnetic field becomes larger in the positive direction. Such a phenomenon of re-entrant superconductivity persists across a temperature span from 400 mK to 600 mK. When the sweep direction of the magnetic field is reversed, the corresponding mapping figure will be mirrored, as shown in Fig. S5b. Furthermore, the re-entrant superconducting states are also observed for magnetic fields in all in-plane directions, corroborated in the supplementary Fig. S6. We label these two distinct superconducting phases as “SC-I” (the superconducting dome obtained before the re-entrance of superconductivity) and “SC-II” (the superconducting dome obtained after the re-entrance of superconductivity) as shown in Fig. 3a. The temperature dependence of the re-entrant behavior is shown in Fig. 3b.

Fig. 3c shows the $\theta$-dependence of the magnetoresistance. It is shown that superconductivity in SC-II is eliminated when the out-of-plane magnetic field angle exceeds 6° while SC-I still exists. This observation underscores the greater fragility of SC-II to out-of-plane magnetic field components. To further display the two superconducting domes SC-I and SC-II of the material, we measure dV/dI as a function of the magnetic field and applied current at different temperatures[70] as shown in Fig.3d. Two superconducting domes SC-I and SC-II are clearly observed. Similar hysteresis and re-entrant behavior are reproduced and Fig. S7&8 show the results for samples #2 and #3.

## Ruling out trivial origins

Although hysteresis in magnetoresistance can be induced by superheating and supercooling effects[29,30] or the flux-trapping effects[26–28] as shown in previous works, the hysteresis loops observed in Fig. 2 are distinct from the hysteresis observed in previous experiments.

In the hysteresis induced by the superheating-supercooling effect, sweeping the magnetic field from 0 T to 1 T (forward sweep) and then gradually reducing the magnetic field from 1T back to 0 T (backward sweep) must result in a higher upper critical field during the forward sweep compared to the backward sweep[29]. In contrast, our observations, as illustrated in Fig. 2f, reveal that during the forward sweep from 0 T to 1 T (indicated by the blue curve), the upper critical field $B_0$ is smaller than the upper critical field $B_{\leftarrow}$ in the backward sweep (indicated by the black curve). The same result is obtained when the magnetic field is swept from 0 T to -1 T and then reduced gradually, as shown in Fig. 2e by the blue and red curves. Such findings exclude the superheating and supercooling mechanism caused by the magnetic field induced local minima of the free energy of the superconductor[71].

In addition, the hysteresis observed in our experiment exhibits behavior starkly distinct from that attributed to the flux-trapping mechanism. In the hysteresis induced by the flux-trapping, a magnetoresistance minimum at finite magnetic field can be observed in the backward-sweeping curve (returning curve)[28,72,73]. In contrast, as shown in Fig. 2 b&c and Fig. S9 in Supplementary Information, in $RbV_3Sb_5$, no magnetoresistance minimum at finite field was observed in the backward-sweeping curve over a wide range of applied currents.

Moreover, we conducted in-plane field-sweep experiments on over ten thin-flake kagome material $CsV_3Sb_5$. Although $RbV_3Sb_5$ and $CsV_3Sb_5$ belong to the same kagome material family and the thin flakes we used have similar thicknesses, no such hysteresis was observed in $CsV_3Sb_5$ below its superconducting transition temperature, as shown in Fig. S10. If the observed hysteresis were due to trivial mechanisms, such as flux trapping, rather than intrinsic properties of the superconducting states, it should also appear in $CsV_3Sb_5$ as well. This result suggests that the superconducting states of $RbV_3Sb_5$ and $CsV_3Sb_5$ are different, and the hysteresis originates from the intrinsic superconducting properties of $RbV_3Sb_5$. It is interesting to note that Knight shift experiments in $CsV_3Sb_5$ suggested spin singlet pairing[74]. However, since our control experiments suggest that the superconducting properties of $RbV_3Sb_5$ are quite different from $CsV_3Sb_5$, the pairing symmetry of $RbV_3Sb_5$ is possibly unconventional.

Furthermore, the re-entrant superconductivity observed in our experiments is different from the re-entrance observed in conventional superconductors[31–33] which relies on vortex pinning induced by applied magnetic field. In contrast to some trivial re-entrant superconductivity observed only in specific directions, i.e. current parallel to magnetic field[32], in $RbV_3Sb_5$, the re-entrant superconductivity appears in all in-plane magnetic field directions with respect to a fixed current direction (as shown in Fig. S6 of Supplementary Information). What's more, the trivial reentrance induced by vortex pinning can only give rise to a symmetric B-T phase diagram[31,33]. This contradicts the asymmetric B-T phase diagram observed experimentally as shown in Fig. 3a.

Therefore, the simultaneous observation of both in-plane magnetic field hysteresis and asymmetric re-entrant superconductivity rules out trivial origins, indicating the need for a novel explanation.

## Possible pairing symmetries

Considering the key observations in experiments reported above, we discuss the possible pairing symmetries in the kagome superconductor $RbV_3Sb_5$. First of all, the observed novel hysteresis in magnetoresistance measurements suggests the spontaneous breaking of time-reversal symmetry. It is noted that the time-reversal symmetry may be broken in the normal states of the kagome family $AV_3Sb_5$ due to loop currents in the charge density wave phase[5,6,13,14]. However, the loop currents would produce out-of-plane magnetization, in contrast to the in-plane magnetic moments observed in this work. Therefore, the in-plane hysteresis is likely caused by the superconducting state itself, rather than the normal state. Moreover, as shown in Fig. 2g, the magnetoresistance hysteresis is weakened by the out-of-plane component of the magnetic field. The reason is as follows: $RbV_3Sb_5$ is a quasi-two-dimensional superconductor[75], where the out-of-plane resistivity is approximately 33 times greater than the in-plane resistivity at low temperatures. In such materials, an out-of-plane magnetic field rapidly suppresses superconductivity due to the orbital effects, which generates many vortices within the superconducting plane. When the applied magnetic field has a sizable out-of-plane component, many vortices will be created, and the vortices will dominate the transport properties of the superconducting films. As the transport is mostly dominated by vortices, the magnetoresistance behaves in a similar way for both the up-sweep and down-sweep fields, leading to the disappearance of hysteresis. A similar behavior is also observed in the heavy-fermion superconductor UCoGe[76], where a slight tilt of the magnetic field toward the c-axis leads to a significant decrease in the upper critical field and the disappearance of hysteresis.

In addition to the pronounced hysteresis, the observation of re-entrant superconductivity as the magnetic field increases suggests that the SC-II phase can be stabilized by the magnetic field. After carefully ruling out trivial origins, the spin-singlet scenario cannot support the field-induced re-entrant superconductivity. This is because the Zeeman effect of the magnetic field tends to suppress the spin-singlet superconducting state by inducing a depairing effect on spin-singlet Cooper pairs, whereas it does not affect the equal-spin pairing states. Therefore, the reappearance of the superconducting state under a larger magnetic field is not consistent with the spin-singlet Cooper pairs scenario. Instead, the observed re-entrance is more consistent with equal-spin pairing, where the spin-magnetic moments of the Cooper pairs couple to the magnetic field, lowering the energy of the superconducting state. Moreover, within the hysteresis loop, the transition of the finite resistance state to the superconducting state upon applying and then removing the current indicates that the finite resistance state is a metastable state. This suggests that there can be different superconducting domains with opposite spin polarizations in the finite resistance state, similar to the case of a ferromagnet with magnetic domains. Furthermore, due to the centrosymmetric point group symmetry, the superconducting pairing should be odd parity due to the spin-triplet pairing[77]. Therefore, the orbital part of the pairing should be p-wave or f-wave in nature. These large number of novel properties observed in the superconducting state place strong constraints on the pairing symmetry of $RbV_3Sb_5$.

Since the superconducting pairing is formed near the Fermi surface, we project the four-orbital (eight-band) model into the manifestly covariant pseudospin basis (MCPB)[65,78,79] by choosing two degenerate bands $(|\vec{k},\alpha\rangle, |\vec{k},-\alpha\rangle)$ near the Fermi surface as the projecting targets. Here, $\vec{k}$ is the crystal momentum and $\alpha$ is the pseudo-spin index (The detail can be found in the Method Section and the Supplementary Information). It is important to note that the original point group symmetry of the material is $D_{6h}$. Before the superconducting transition happens, the nematic order parameter comes in[10,14] and reduces the point group symmetry to $D_{2h}$. Therefore, the superconducting order parameters can be constructed by using basis functions belonging to irreducible representations of the point group $D_{2h}$. As mentioned above, this pairing state must possess net in-plane spin polarization. The possible superconducting order parameters represented by the $\vec{d}$-vectors are listed in Table I. Importantly, both of these are nodal topological p-wave, spin-triplet, time-reversal-symmetry-breaking states. As shown in the Supplementary Information, both of these states possess Majorana flat bands at the sample boundaries. In the following, we choose the $B_{3u}$ representation without loss of generality.

| Representations | Order parameters |
|---|---|
| $B_{2u}$ | $\vec{d}_{2u} = (\sin(k_2 - k_1) + \sin(k_2))(\cos(\chi)\cos(\xi), \sin(\chi)\cos(\xi), i\sin(\xi))$ |
| $B_{3u}$ | $\vec{d}_{3u} = \sin(k_1)(\cos(\chi)\cos(\xi), \sin(\chi)\cos(\xi), i\sin(\xi))$ |

**Table I Possible order parameters with time-reversal symmetry breaking.** Here $k_1 = k_x$, $k_2 = \frac{1}{2}k_x + \frac{\sqrt{3}}{2}k_y$.

The matrix representation of the superconducting gap in the MCPB is $\hat{\Delta} = (\vec{d}\cdot\vec{\sigma})i\sigma_2$, where the $\sigma_i(i = 1,2,3)$ are the Pauli matrices. The nodes of these two order parameters belonging to different representations are located on x-axis and y-axis, respectively. The parameter $\xi$ controls the amplitude of the spin polarization: $1 - \eta = 1 - |\cos(2\xi)^2 - \sin(2\xi)^2|$, which is determined by the specific form of the interactions in this material, while the parameter $\chi$

controls the spin-polarization direction of the Cooper pairs. The orientation of the spin polarization $\vec{p}$ of the Cooper pair is given by: $\vec{p} = i\vec{d} \times \vec{d}^{*} \propto (\sin(\chi), -\cos(\chi), 0)$, which is coupled with the effective magnetic field $\vec{g}_{\vec{k}}$ in the MCPB in the form of $-\vec{g}_{\vec{k}} \cdot \vec{p}$. Therefore, the direction of the spin polarization can be pinned by the applied magnetic field (See Supplementary Information for detail). When the magnetic field changes its sign, superconducting domains with opposite spin-polarization become energetically favorable and can be created.

## A possible mechanism for the hysteresis, current heating and cooling experiments, and re-entrant superconductivity

Here, we provide a possible, schematic superconducting domain picture to explain the hysteresis in resistance, the current heating and cooling experiments as well as the reentrant superconductivity observed in $RbV_3Sb_5$. As discussed in last section, the system has spin-polarized triplet pairing components which can form the superconducting domains[80,81] where Cooper pairs with net spin-polarization act like miniature magnets with different polarization directions, similar to ferromagnetic domains. At a low temperature for example (as denoted by $T_1$ in Fig. 4), when the magnetic field sweeps from -1 T to 1 T, the system initially enters into a superconducting state dominated by negative domains. The negative domains are formed by spin-polarized Cooper pairs with magnetic moments pointing to the negative direction as indicated by blue arrows in Fig. 4. In other words, the spin-polarized Cooper pairs can form domains and the external magnetic field aligns the magnetization direction of the domains. When the magnetic field changes sign, superconducting domains with positive magnetization direction (positive domains) emerge. As the magnetic field further increases in the positive direction, positive domains become dominant, but the negative domains can still coexist. Therefore, at a positive magnetic field, we expect that the system enters into a metastable state characterized by the coexistence of the positive and negative domains as shown in Fig. 4. On the other hand, if the magnetic field is swept from 1 T to the same positive magnetic field, the domains are dominantly positive and the superconducting state will have a lower free energy. This provides a possible explanation for the hysteretic behavior of the resistance. Moreover, at a fixed positive magnetic field ~ 0.27 T (Fig. 2h), we can use a large current to heat the metastable finite-resistance state with both positive and negative domains and then remove the current under a fixed magnetic field, the system can transition to the superconducting ground state, where superconducting domains predominantly align with the applied magnetic field. This current heating and cooling experiment strongly supports the superconducting domain picture of our theory which suggests that the SC-II state is a metastable state containing superconducting domains with different spin polarizations. In addition, in Fig. S12, three curves of distinct colors correspond to different initial magnetic fields (1 T for the black curve, 0 T for the blue curve, and -1 T for the red curve) at the onset of the field sweeping. These disparate trajectories, as illustrated in Fig. S12, indicate that the magnetic response of the superconducting state in $RbV_3Sb_5$ is determined by the initial magnetic field, thereby further supporting the domain picture.

At a higher temperature (as denoted by $T_2$ in Fig. 4), when the magnetic field sweeps from -1 T to 1 T, the system is initially in a superconducting state with negative domains. When the magnetic field changes sign from negative to positive, the negative domains shrink and the positive domains start to form as in the low temperature cases. However, due to the thermal effect, these domains are too small to be connected spatially to form a superconducting state. As the magnetic field increases, the positive domains become more energetically favorable to form. The enlargement of positive domains allows percolation of superconducting domains to give rise to superconducting state. This percolation picture gives a mechanism for the re-entrance of superconductivity at around 400 mK as observed in the experiment. Therefore, the nodal p-wave superconductivity belonging to the $B_{2u}$ or the $B_{3u}$ representation of the $D_{2h}$ point group, which enables Cooper pairs exhibiting net spin-polarization, can provide a reasonable explanation of all the observed novel phenomena in $RbV_3Sb_5$.

## Conclusion and Perspective

In summary, we report the experimental evidence of intrinsic time-reversal symmetry-breaking p-wave superconductivity in $RbV_3Sb_5$. The observed hysteresis in the superconducting state, the re-entrance of superconductivity, the current heating and cooling experiments, and the two-fold symmetric in-plane critical field, impose strong constraints on the possible pairing state of $RbV_3Sb_5$. Our symmetry analysis suggests that the pairing state is consistent with a spin-polarized p-wave pairing state, characteristic of a nodal topological superconducting state. This state supports Majorana flat bands[41–48] on the edge of the samples. It was proposed that the large number of Majorana zero energy modes associated with the Majorana flat bands will induce a high zero bias peak in the tunneling experiment[82–85]. Our work points out that $RbV_3Sb_5$ is possibly a novel platform for studying intrinsic topological superconductivity.

# Methods

## Sample fabrication

The thin films of $RbV_3Sb_5$ were exfoliated using polydimethylsiloxane (PDMS) from the bulk crystal onto silicon substrates with a 285 nm oxide layer. The fabrication of pre-patterned metal electrodes was achieved using the standard electron-beam lithography (EBL) method, followed by deposition of a Ti/Au (10 nm/50 nm) through electron-beam evaporation. Then, the device was encapsulated with hBN to prevent any degradation due to air atmosphere. All device fabrication procedures mentioned above were performed in a controlled environment within a nitrogen-filled glovebox with oxygen and water levels below 0.01 ppm to minimize any potential degradation of the samples.

## Quantum transport measurements

The samples were measured in an Oxford dilution refrigerator Triton XL1000 equipped with a vector magnet. This setup allowed the samples to remain stationary while a real three-dimensional vector magnetic field was applied. Such a technique greatly improves the precision of the magnetic field angle-deviation measurements, effectively circumventing the issues of return differences and heat generation commonly associated with mechanical rotators.

The dV/dI measurements were conducted with a rather small alternative current (i.e. 100 nA or 10 nA) while applying a much larger direct current (i.e. 10 μA). To optimize the signal-to-noise ratio in these experiments, a standard lock-in technique was employed, along with the implementation of essential low-temperature filters.

## Conducting the strict in-plane magnetic field measurements with the 3D vector magnet

1. Initially, we securely mounted the sample within the dilution refrigerator. Despite our meticulous efforts to align the sample with the magnetic field axis, a minor discrepancy persisted.
2. In response to this, our subsequent step involved the calibration of the vector magnetic field. What sets our approach apart from the traditional mechanical rotational system is that we did not need to physically rotate the sample plane, we could fine-tune the value of the vector magnetic field, which would greatly enhance the angle accuracy.
3. During the calibration process, we initially conducted angle-dependent measurements of the critical magnetic field of the sample in both the XZ and YZ planes. Given that our sample plane was mounted in the XY plane, these measurements allowed us to determine the angle deviation from the sample plane. With the calibration of the XZ-plane and YZ-plane, then the new coordinates are achieved, with strict alignment of the vector magnetic field within the XY plane. It's worth noting that our device substrate consisted of a rigid silicon wafer with a 285 nm $SiO_2$ layer. Through this dual-axis calibration process, we can confidently assert that the calibrated XY plane remains strictly within the sample plane, with an error margin of less than 0.1°.

**Ruling out trivial origins of two-fold symmetric superconductivity**

It has been suggested that the anisotropy may arise from current-induced vortex motion. Such a scenario was ruled out by different configurations in our previous paper[86]. The two-fold symmetric axis is independent on the current direction. Specifically speaking, in both cases that the current is applied in-plane with $\phi$=90° and $\phi$=0°, the in-plane azimuthal angular dependence of the magnetoresistance exhibited the same twofold symmetry pattern, irrespective of the current direction. Thus, the two-fold symmetry is independent of the current direction, ruling out the possible vortex origin.

**Manifestly covariant pseudospin basis (MCPB)**

Due to the existence of the inversion symmetry, though there is SOPC[65,66], one can always define a pseudospin basis, in which the orbital and pseudospin degrees of freedom are decoupled. On the other hand, since the superconducting pairing happens near the Fermi surface, we can project the four-orbital (eight-band) model into a two degenerate band basis with different pseudospins, which is named as the manifestly covariant pseudospin basis (MCPB) [65,78,79]. In this basis, the magnetic field couples to the spin in this form:

$$H_Z = \mu_B \sum_{\vec{k},\alpha,\beta,i,j} c^{\dagger}_{\vec{k}\alpha} B_i a_{ij}(\vec{k}) \sigma_j^{\alpha\beta} c_{\vec{k}\beta},$$

where $\mu_B$ means the Bohr magnetron, $\alpha, \beta$ label the pseudospin of the fermion in band basis, $\sigma_i (i = 1,2,3)$ are Pauli metrices. Here, $a_{ij}(\vec{k}) (i, j = 1,2,3)$ come from the projection to MCPB (See Supplementary Information for detail), which change the original magnetic field $B_i$ into a k-dependent effective magnetic field $\vec{g}_{\vec{k}} = \mu_B \overleftrightarrow{a}_{\vec{k}} \vec{B}$. Including the pairing order parameter, the total Hamiltonian of the superconducting state can be written as:

$$H_{total} = \sum_{\vec{k},\alpha} \varepsilon_{\vec{k}} c^{\dagger}_{\vec{k}\alpha} c_{\vec{k}\alpha} + \sum_{\vec{k}\alpha\beta} (c^{\dagger}_{\vec{k}\alpha} [(\vec{d} \cdot \vec{\sigma}) i\sigma_2]_{\alpha\beta} c^{\dagger}_{-\vec{k}\beta} + h.c.) + H_z,$$

where the $\vec{d}$-vector is the possible order parameter of the odd-parity superconductivity which is show in Table I. With the total Hamiltonian $H_{total}$, one can calculate the topological properties and the magnetic responses conveniently. Importantly, this Hamiltonian describes a nodal topological p-wave superconductor which possesses the Majorana flat bands on the edges of the samples (See Fig. S13 in Supplementary Information for the result of the open boundary calculations).

## Data availability

The data that support the plots within this paper and other related findings are available from the corresponding author upon reasonable request.

## Acknowledgments

This work was supported by the National Key Research and Development Program of China (Grants No. 2020YFA0309300, No. 2022YFA1403700), the National Natural Science Foundation of China (Grants No. 12074162, No. 12004158, No. 91964201), the Key-Area Research and Development Program of Guangdong Province (Grant No. 2018B030327001), Guangdong Provincial Key Laboratory (Grant No.2019B121203002), Guangdong Basic and Applied Basic Research Foundation (Grant No. 2022B1515130005) and Guangdong province 2020KCXTD001.

## Author contributions

B.-C.L. proposed and designed the experiment. S. W. and B.-C.L. did the quantum transport experiments with J.-J. Z.'s software support. S.W. and J.-Z.F. fabricated and characterized samples with necessary assistance of J.-P.P., J.-J.Y. and Z.Z. J.-J.Y. and J.-P.P. did the AFM experiments. J.L. and Xuewen F. did the HRTEM experiments. N. K. and X.-S. W. helped with the understanding of experiment data. Xilin F., Z.-T.S., and K.T.L. proposed the theory. B.-C.L. and D.Y. supervised the whole project. B.-C.L., Xilin F., and K. T. L. wrote the manuscript with inputs from all authors.

## Competing interests

The authors declare that they have no competing interests.

## Figures and Tables

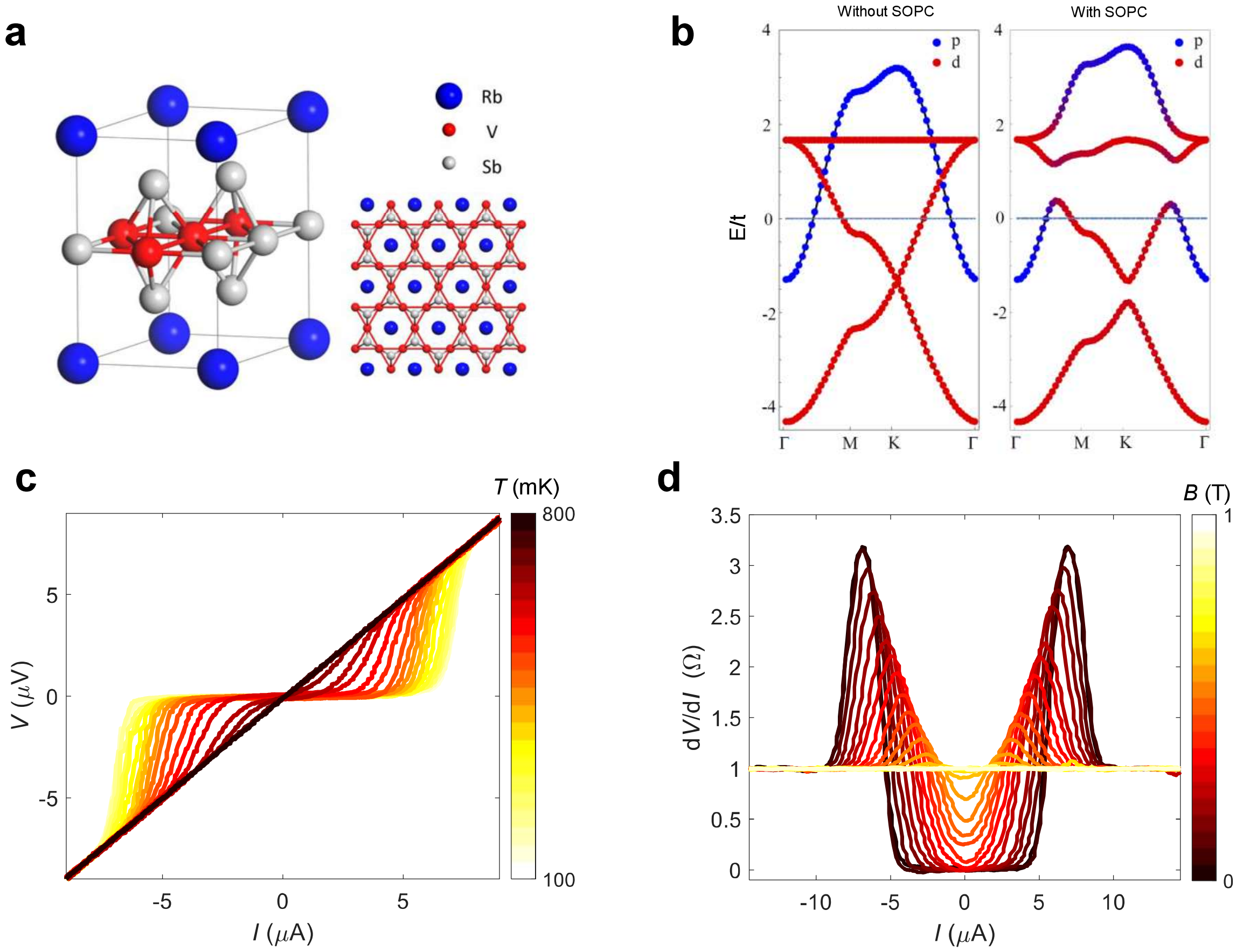

**Fig. 1 The crystal structure and superconducting behavior of $RbV_3Sb_5$. a,** The crystal structure of the kagome material $RbV_3Sb_5$. **b,** The band structure of the minimal four-band model describing the in-plane properties of the kagome superconductor without (left) and with (right) spin-orbital-parity coupling (SOPC) effect, respectively. The SOPC leads to the mixing of the d-orbital belonging to V atoms and the p-orbital belonging to Sb atoms near the Fermi energy. Here, t represents the hopping strength between the nearest-neighbor V atoms. **c,** The current-voltage curve at different temperatures. **d,** The differential resistance dV/dI versus bias current I as a function of the magnetic field at 150 mK shows the typical superconducting behavior.

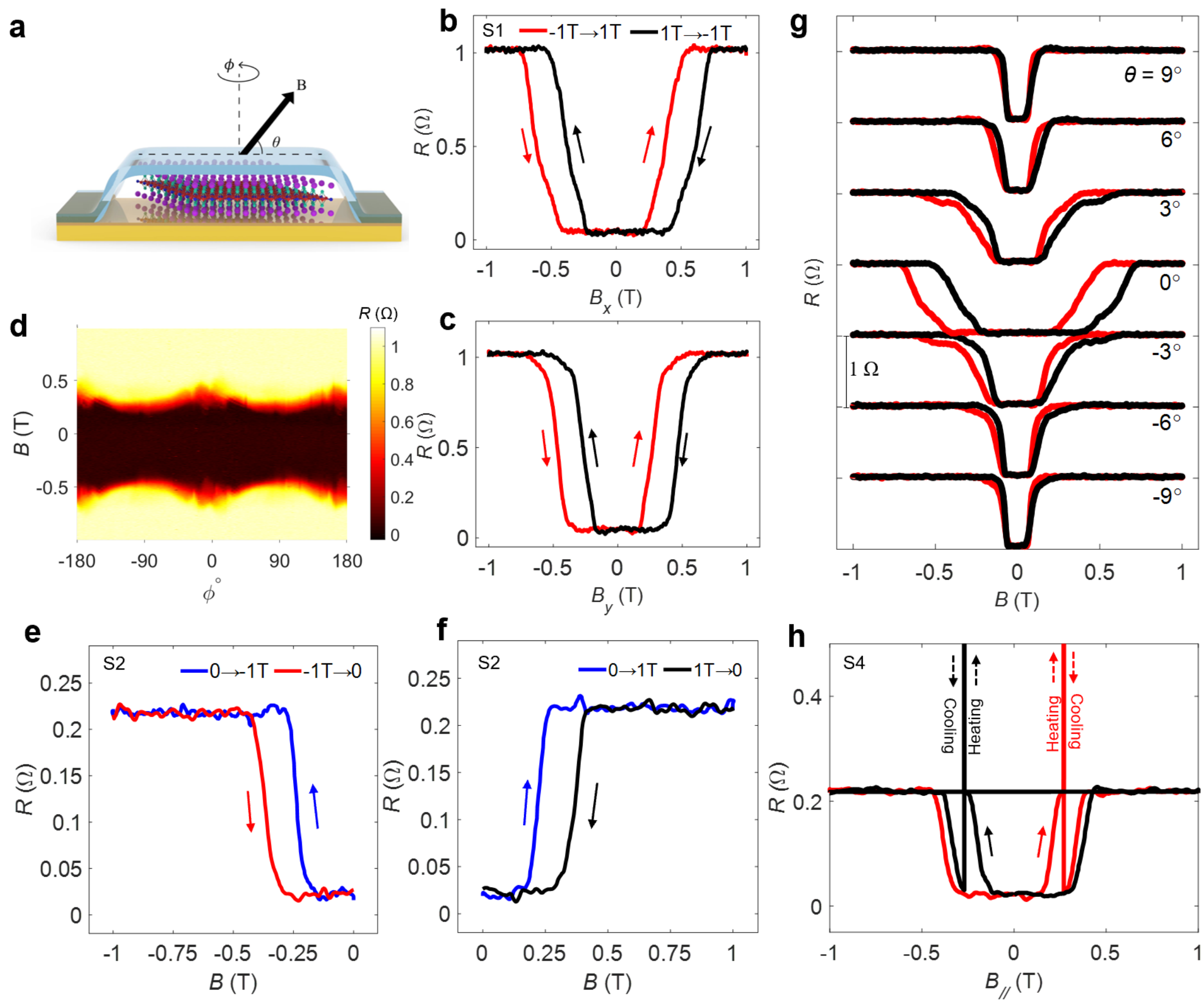

**Fig. 2 The hysteretic behavior of the superconducting states. a,** The schematics of the device. **b&c,** The in-plane magnetic field dependence of the magnetoresistance shows the hysteretic behavior along x-axis and y-axis. The temperature is 150 mK. Notably, the critical field values—defined by the magnitude of the magnetic field at which the resistance reaches 50% of the normal state resistance at the positive magnetic field—during forward (denoted as $B_{\rightarrow}$) and backward ($B_{\leftarrow}$) sweeps exhibit $B_{\rightarrow} < B_{\leftarrow}$. **d,** The mapping plot of the magnetoresistance versus the magnetic field and its direction. **e&f,** The magnetoresistance curve measured during the field-sweeping process of sample S2: the two blue curves, positioned on either side of the B = 0 axis, representing the cases where the field is swept from 0 T to 1 T and from 0 T to -1 T, respectively. The red and black curves represent the magnetoresistance measured when the field is swept from -1 T to 0 T and 1 T to 0 T, respectively. **g**, The out-of-plane angle dependence of the magnetoresistance. **h**, The current heating and cooling process of the hysteresis in resistance as a function of the in-plane magnetic field at a temperature of 230 mK. A large current is applied at a finite resistance state to heat up the sample. After the current is removed, the sample transits to a superconducting state.

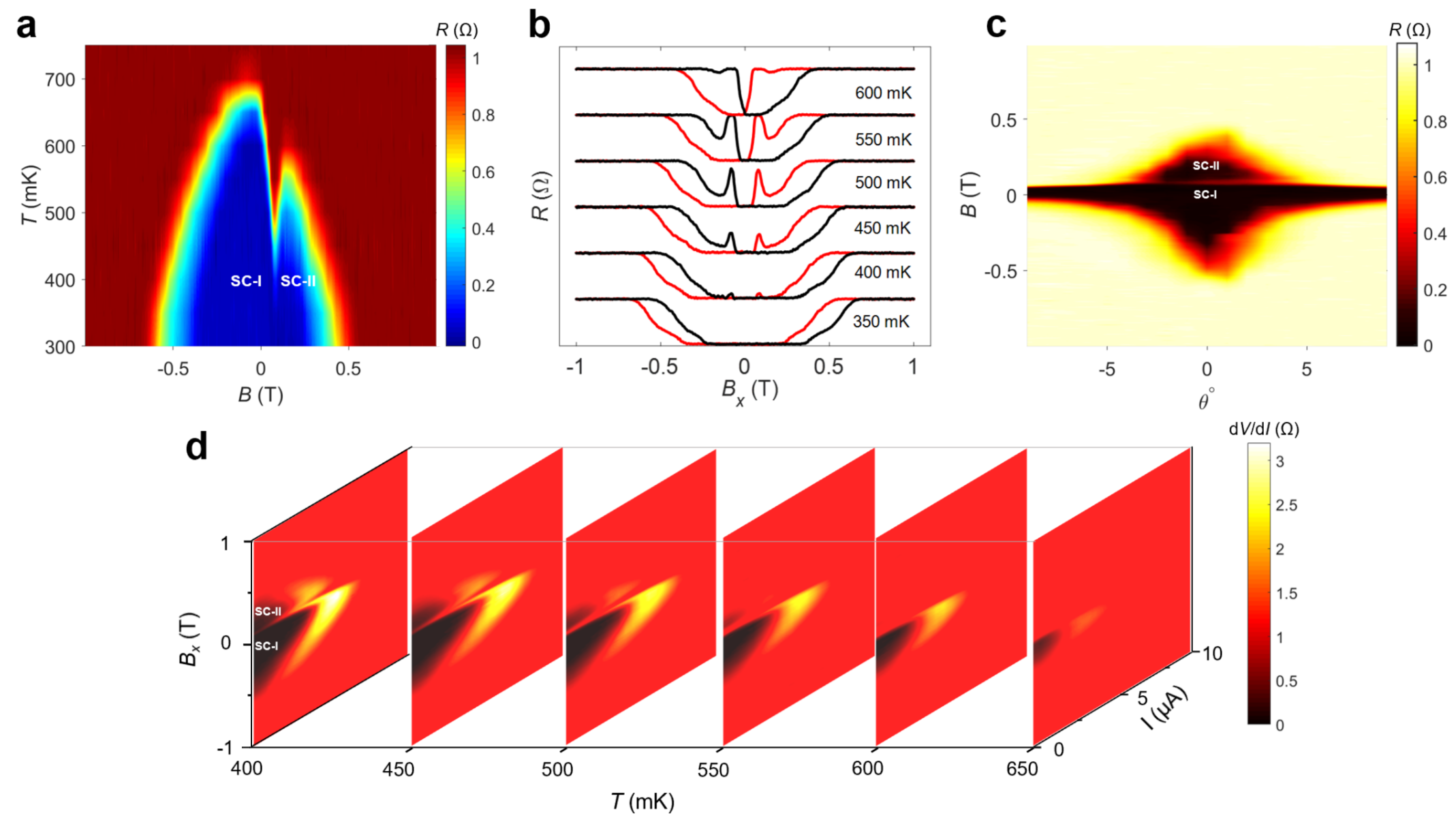

**Fig. 3 The re-entrant superconductivity. a,** The temperature dependence of the magnetoresistance when the magnetic field sweeps from -1 T to 1 T. With temperature increasing, the critical magnetic field gradually decreases. However, a re-entrant superconductivity occurs at around 400 mK. Here we label the superconducting dome obtained before the re-entrance of superconductivity as SC-I and the superconducting dome obtained after the re-entrance of superconductivity as SC-II. **b,** The waterfall plot of the resistance at different temperatures. It is evident that the re-entrant superconductivity starts at 400 mK. The red curves denote a forward sweep, while the black curves represent a backward sweep. **c,** The out-of-plane angle dependence of two superconducting states at a temperature ~ 400 mK. **d,** The temperature dependence of the mapping plot of the two superconducting states versus the magnetic field and bias current.

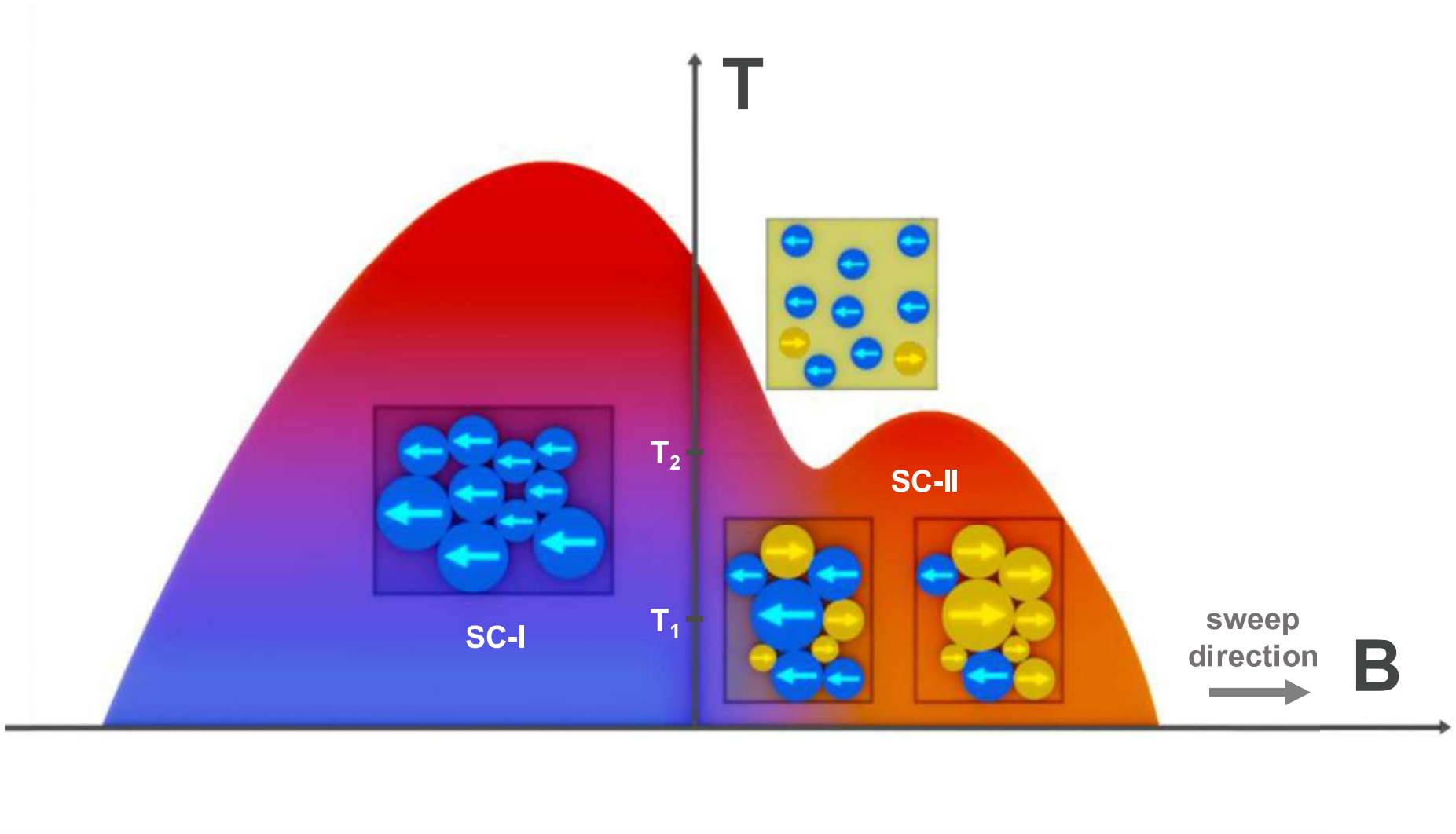

**Fig. 4 The schematics of the mechanism of spin-polarized p-wave superconductivity.** Here 'T' and 'B' denote the temperature and the magnetic field, respectively. As the magnetic field sweeps from -1 T to 1 T, the system initially enters into a superconducting state dominated by 'negative' equal spin pairing domains, indicated by blue arrows in the first dome of superconductivity, namely SC-I. At a low-temperature $T_1$, when the magnetic field changes sign, 'positive' domains, represented by yellow arrows, begin to proliferate. This coexistence of negative and positive superconducting domains results in the formation of a metastable state, SC-II, with higher free energy than SC-I. As a consequence, it reduces the transition temperature and the upper critical field, leading to an asymmetry between the two superconducting domes. At a high-temperature $T_2$, when the magnetic field changes sign, the 'negative' domains become disconnected, and the system enters a finite-resistance state. At higher magnetic fields, the positive domains become more energetically favorable to form, and the proliferation and percolation of 'positive' domains lead to the second dome of superconductivity.

Supplementary Materials for

# Signatures of spin-polarized p-wave superconductivity in the kagome superconductor $RbV_3Sb_5$

*Corresponding Email:
phlaw@ust.hk
linbc@sustech.edu.cn
yudp@sustech.edu.cn

**This PDF file includes:**

**Section 1 The experiments**

**Section 2 The theory of spin-polarized p-wave superconductivity**

## Section 1 The experiments

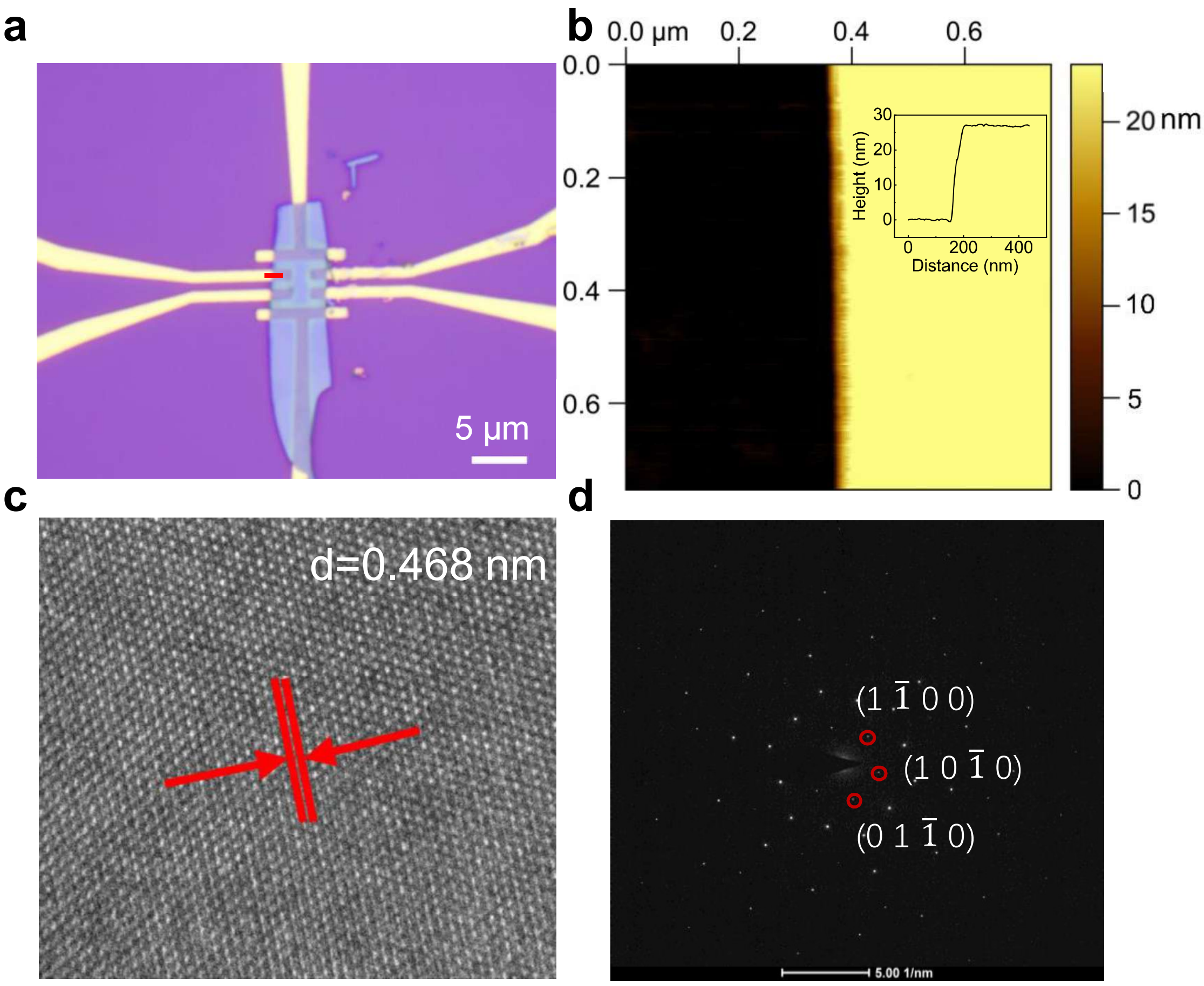

**Fig. S1 Crystal structure and characterizations of $RbV_3Sb_5$. a,** The optical image of the device. **b,** The corresponding atomic force microscope (AFM) measurement of the sample, which shows the flake is around 27.6 nm. **c,** The high-resolution transmission electron microscope (HRTEM) image of $RbV_3Sb_5$, indicating that the $RbV_3Sb_5$ is single-crystalline with lattice fringes of 4.68 Å corresponding to the $(10\bar{1}0)$ plane, which is further confirmed by the selected-area electron diffraction (SAED) pattern given in **d**.

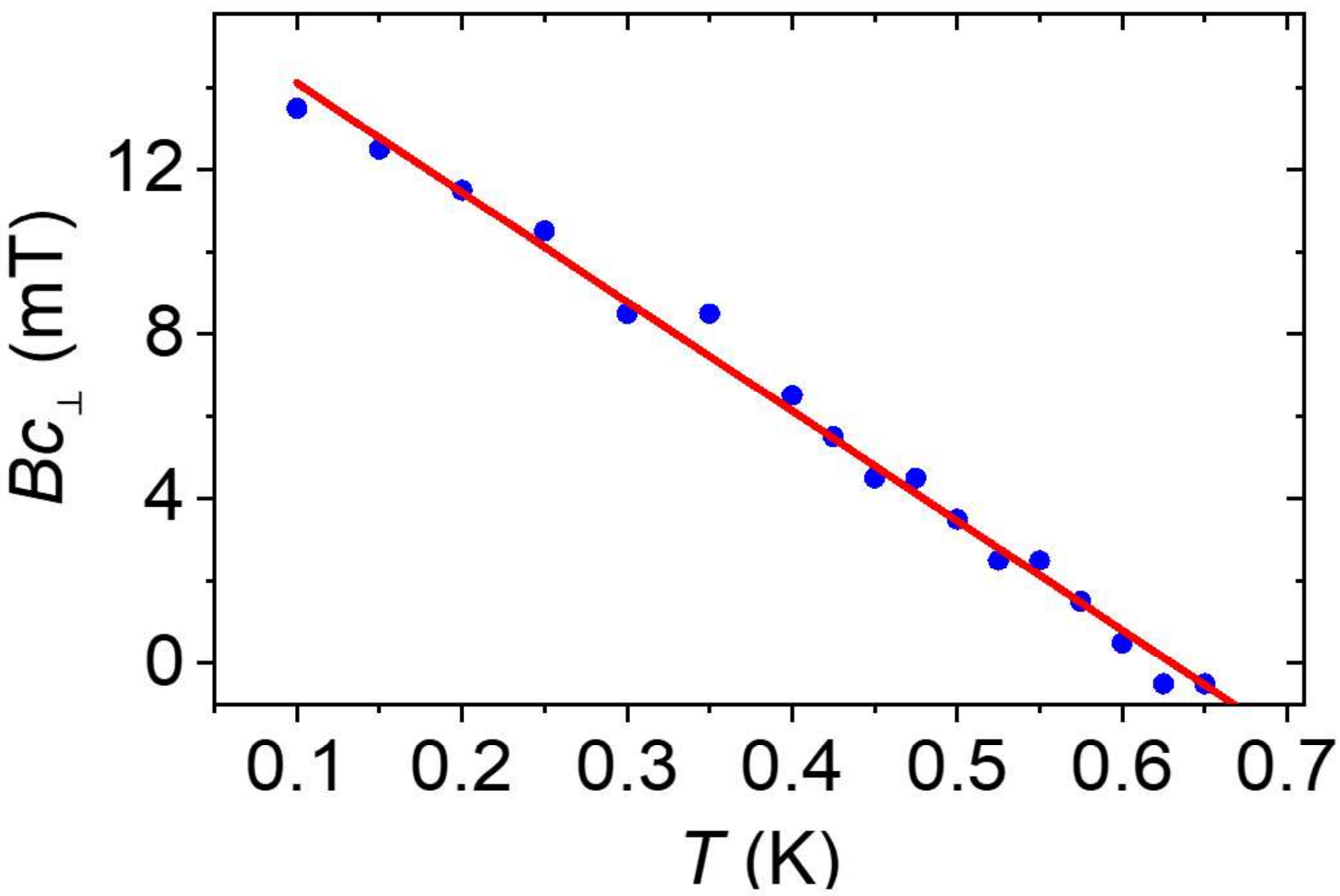

**Fig. S2 The temperature dependence of the out-of-plane critical magnetic field.** The blue dots denote the out-of-plane critical magnetic field, while the red line shows the linear fitting according to the linearized Ginzburg–Landau (GL) expression $Bc_\perp = \Phi_0/2\pi\xi(0)^2\,(1 - T/T_C)$, where $\Phi_0$ is the magnetic flux quantum and $\xi(0)$ is the zero-temperature GL coherence length.

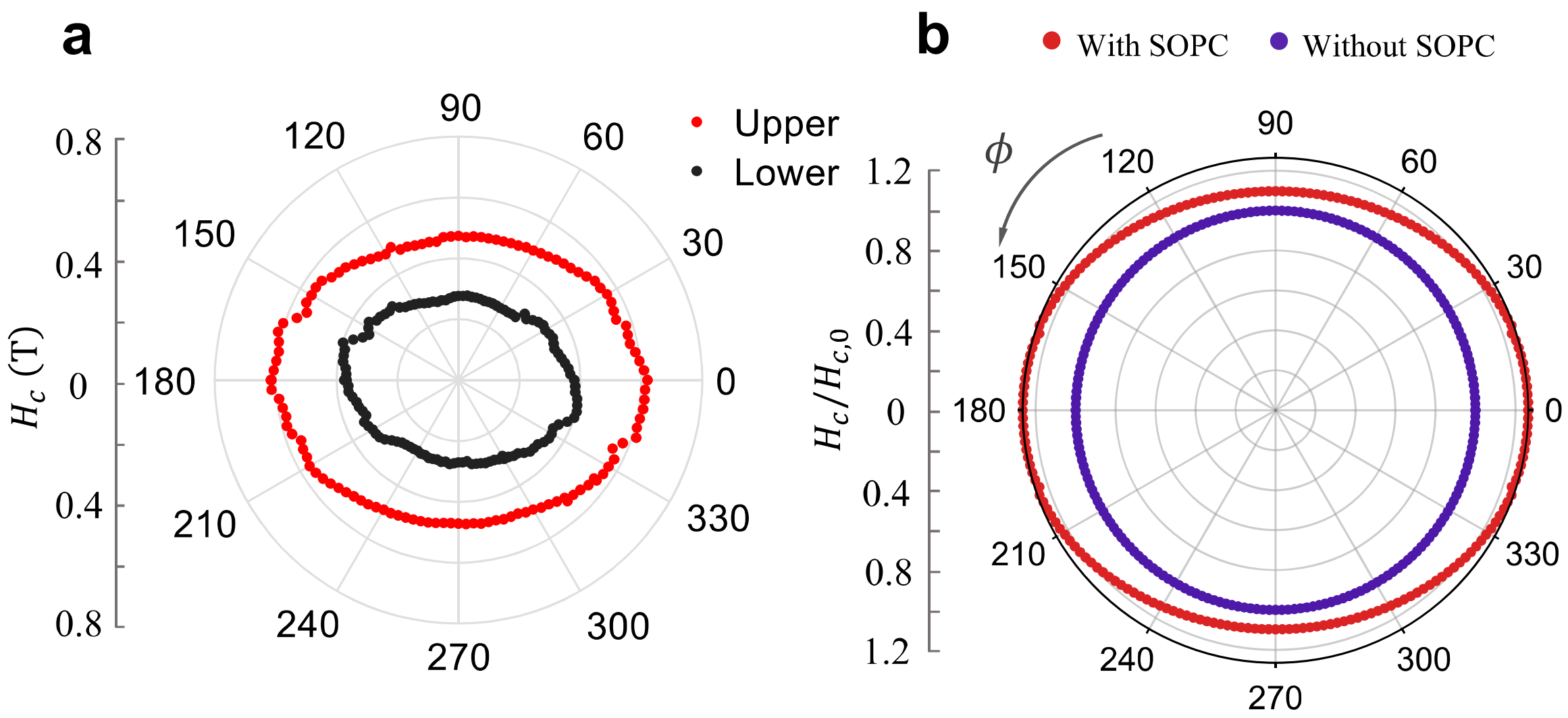

**Fig. S3 The two-fold symmetry of the upper critical field. a**, The two-fold symmetry of the upper and lower critical magnetic fields, extracted from Fig. 2d. Here the upper and lower magnetic fields are extracted from 50% of the normal state resistance. The temperature is 150 mK. **b**, Calculated angular dependence of the upper critical magnetic field for the case with and without spin-orbit-parity coupling (SOPC) in the single-layer limit. With SOPC in the single-layer limit, there is a clear two-fold symmetry in the upper critical field $H_c$.

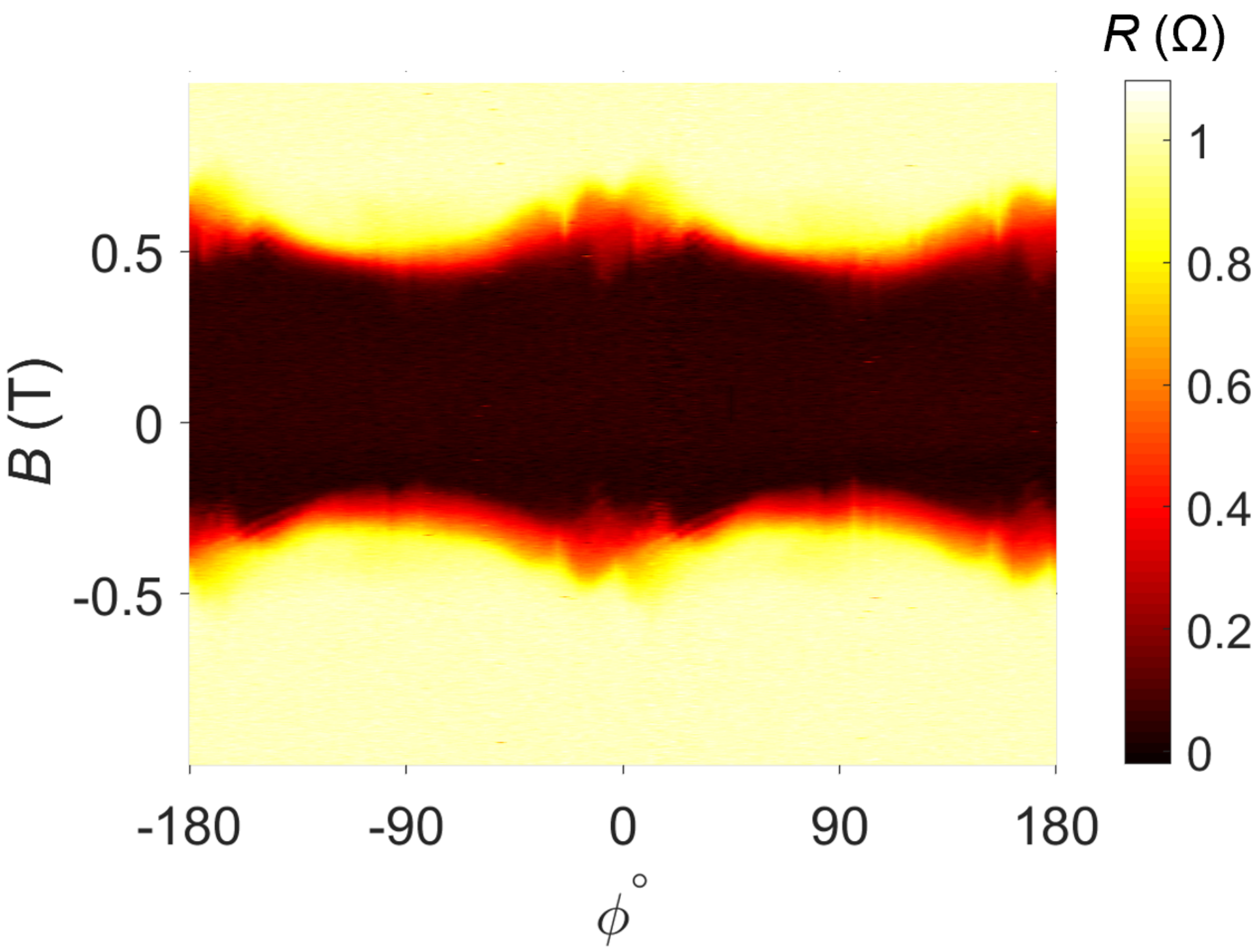

**Fig. S4 The in-plane magneto-resistance with the magnetic field sweeping from 1 T to -1 T.** Correspondingly, the hysteresis is mirrored as the magnetic field sweeps from -1 T to 1 T. The temperature is 150 mK.

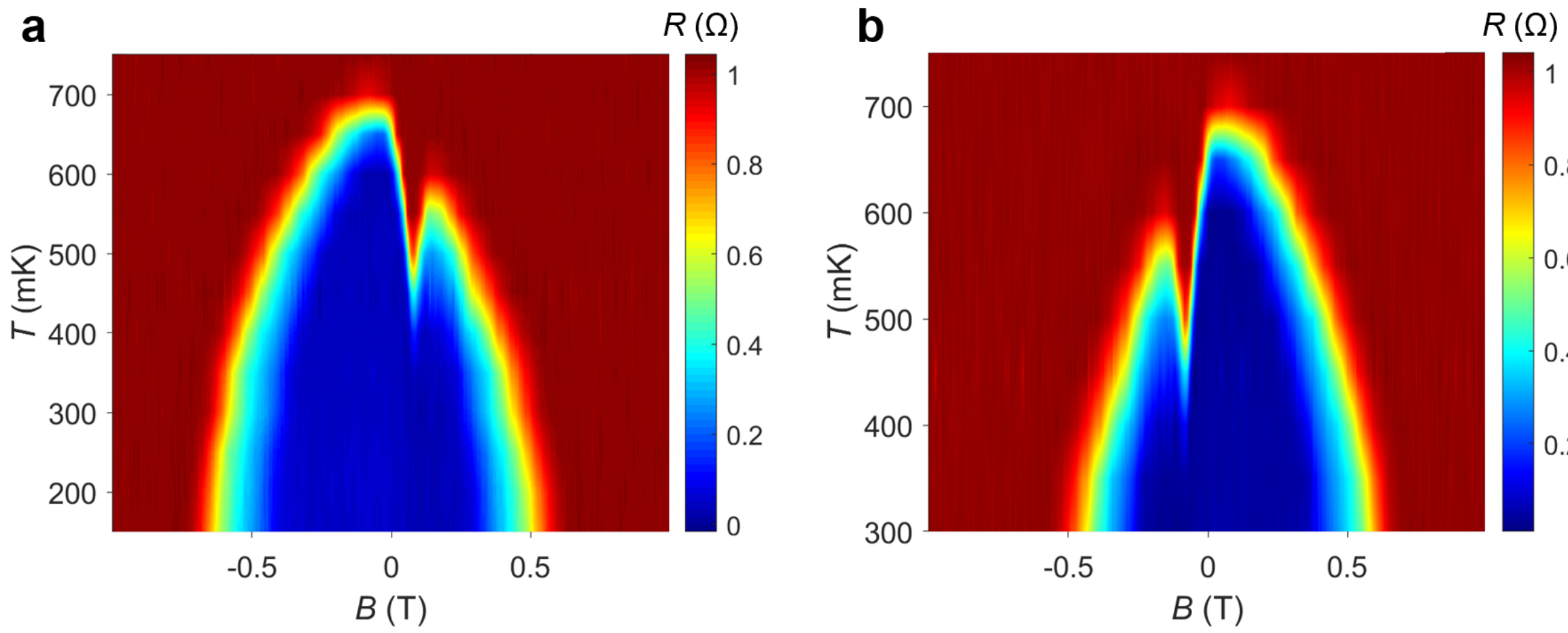

**Fig. S5 The re-entrant superconductivity. a,** The magneto-resistance at all temperatures. **b,** When the magnetic field sweeps from 1 T to -1 T (backward), the figure would be mirrored to Fig. 3 in the main text.

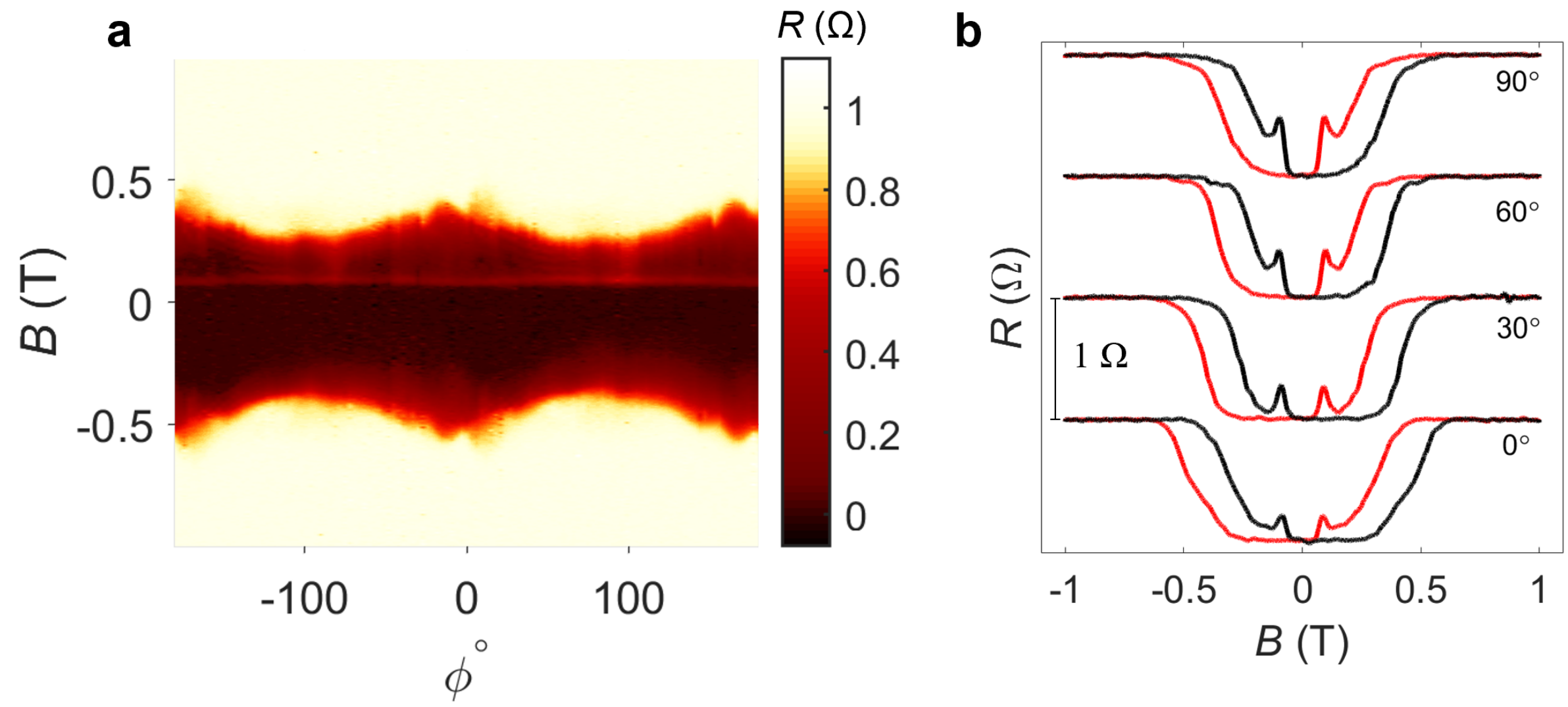

**Fig. S6 The re-entrant superconductivity under different in-plane magnetic fields with angles ranging from -180° to 180°. a**, The mapping plot of the re-entrant superconductivity. **b**, The corresponding waterfall plot of the re-entrant superconductivity at different in-plane azimuthal angles. The temperature is 450 mK.

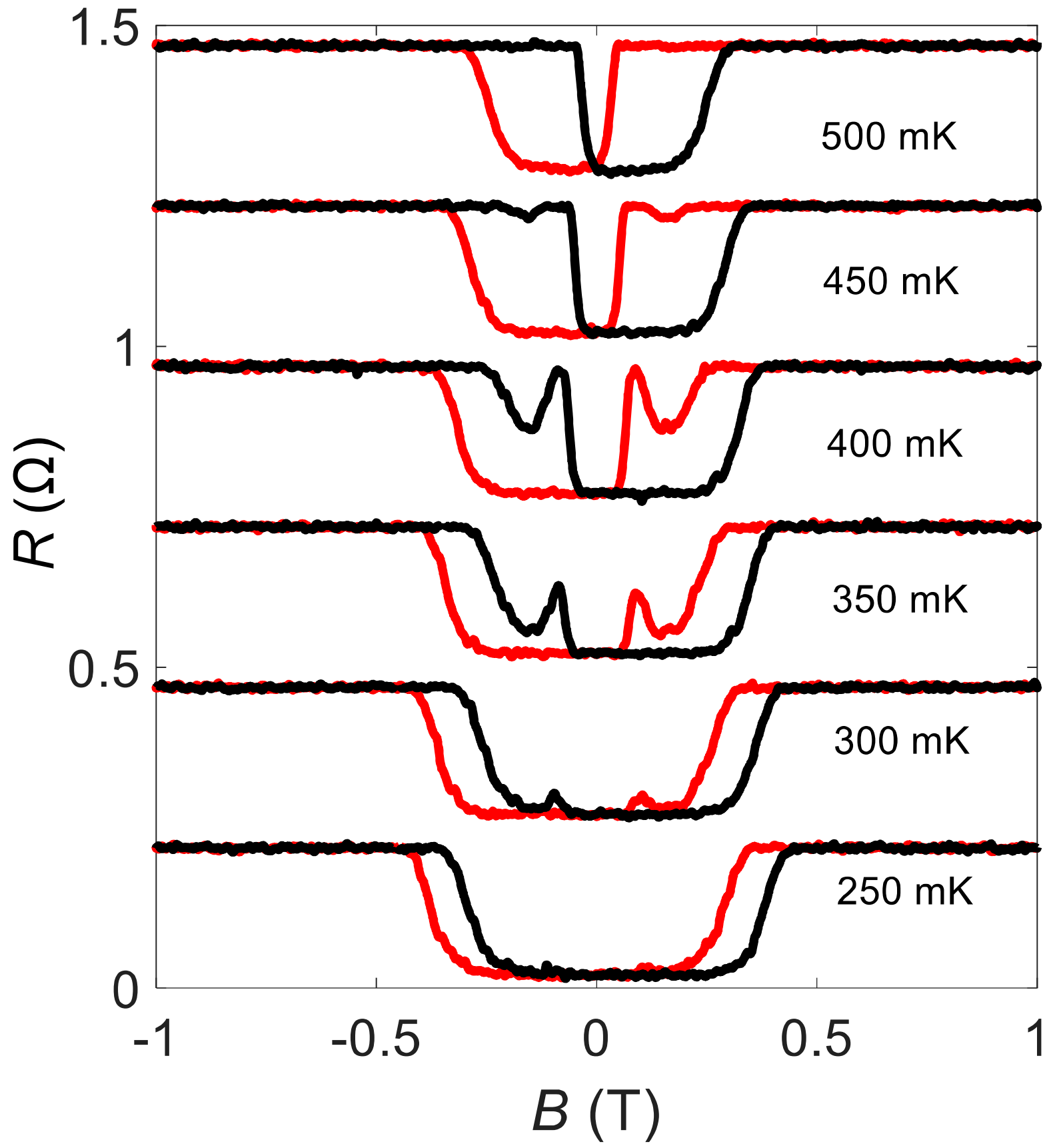

**Fig. S7** The hysteresis and re-entrant superconductivity observed in another sample #2. The red and black lines denote the forward and backward sweep of the magnetic fields, respectively.

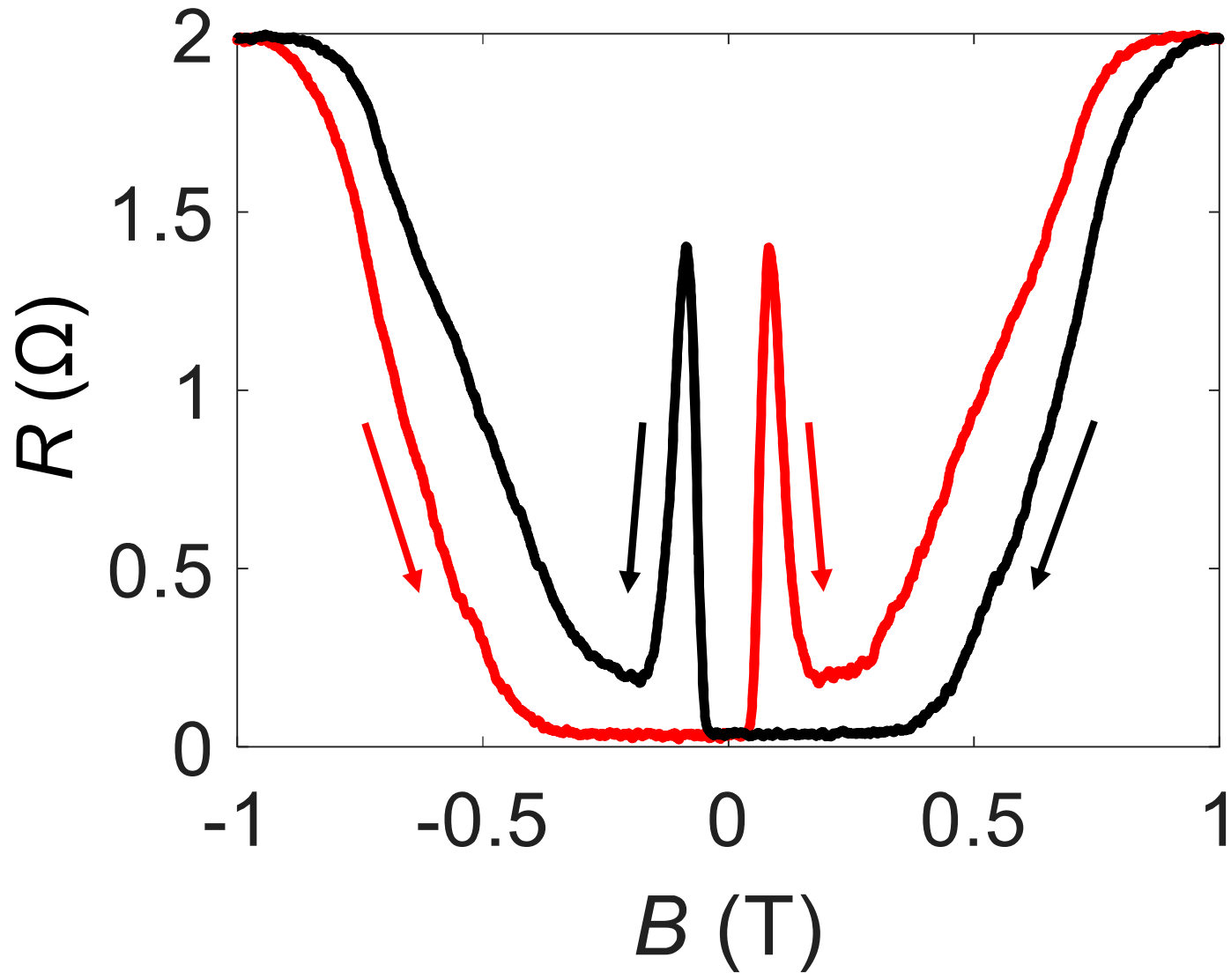

**Fig. S8** The hysteresis and re-entrant superconductivity in another sample #3. The temperature is 400 mK.

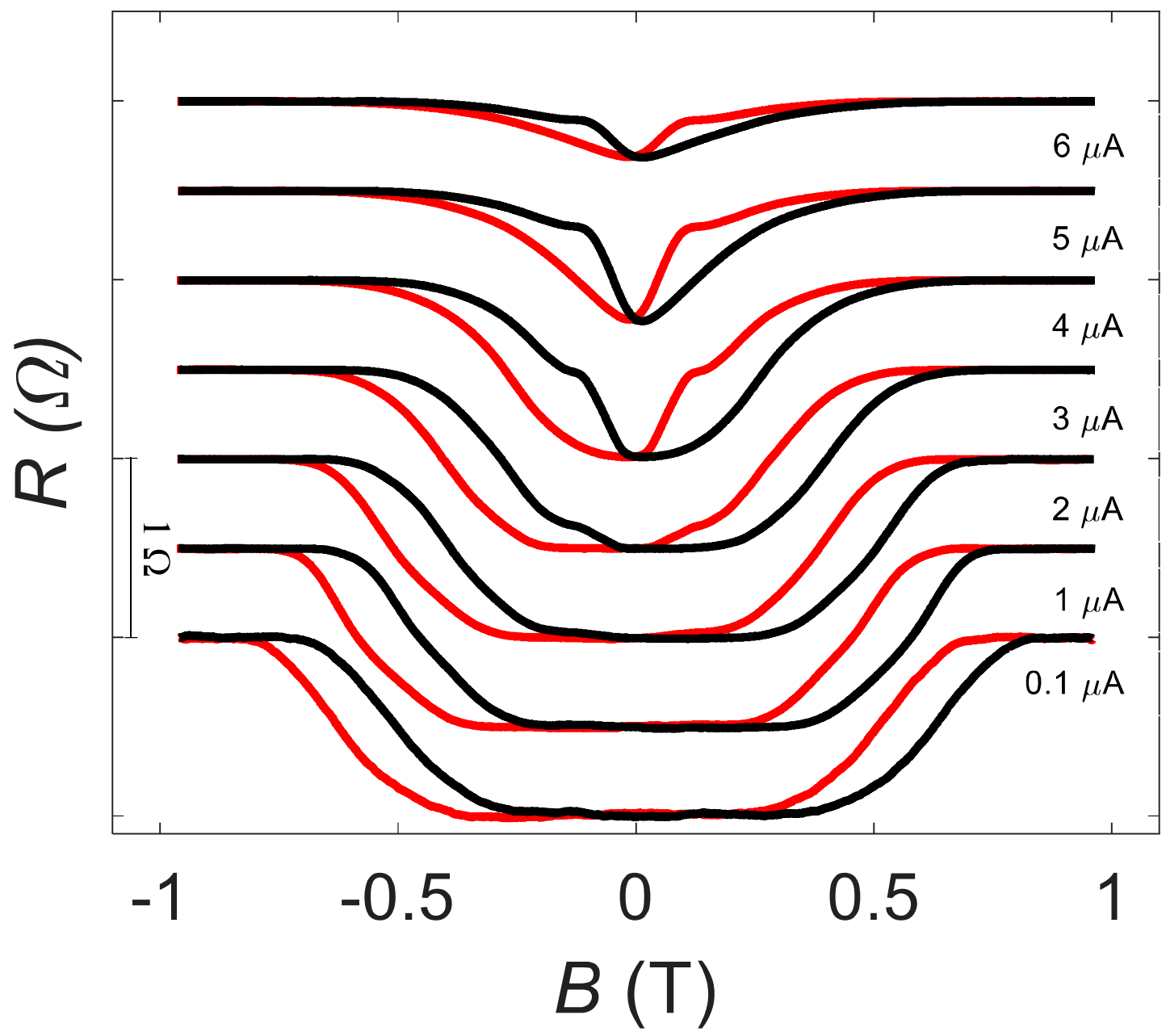

**Fig. S9.** The current-dependent filed-sweeping experiments of the thin-flake sample at a temperature of T = 150 mK.

## Note 1: Discussion on differences between $RbV_3Sb_5$ and $CsV_3Sb_5$

1. **Difference between bulk $RbV_3Sb_5$ and $CsV_3Sb_5$.** At first glance, $RbV_3Sb_5$ and $CsV_3Sb_5$, as members of the same kagome superconductor family, might be expected to exhibit similar properties. However, experimental evidence reveals significant differences in both their normal and superconducting states. For instance, ARPES measurements indicate distinct charge density wave (CDW) structures in the two materials (Nature Materials 22, 186-193(2023), PRB 106, L241106 (2022)), which could influence their superconducting properties, given the competitive relationship between CDW and superconductivity. Additionally, μ-SR experiments suggest that $RbV_3Sb_5$ is more likely to be a nodal superconductor while $CsV_3Sb_5$ appears to be nodeless (Nature Communications 14, 153(2023)). The presence of residual in-gap states in $RbV_3Sb_5$ further hints at unconventional pairing mechanisms. Consequently, the superconductivity in $CsV_3Sb_5$ may exhibit different behavior from that in $RbV_3Sb_5$.
2. **Theoretical perspective on $CsV_3Sb_5$:** While some experiments suggest that $CsV_3Sb_5$ may exhibit spin-singlet or even s-wave superconductivity, theoretical studies caution against oversimplifying its nature. For example, recent works (e.g. Physical Review B, 110, 144516 (2024), Physical Review B 108, 144508 (2023)) highlight that unconventional superconductivity on the kagome lattice is deceptive in the sense that its properties may appear similar to conventional non-sign-changing superconductivity. Specifically, while Knight shift suppression and the observation of a Hebel-Slichter peak below $T_c$ in the spin-lattice relaxation rate measured by NMR experiments were interpreted as evidence for s-wave spin-singlet superconductivity (Chin. Phys. Lett. 38 077402 (2021)), it is proposed that such features can also arise in unconventional kagome superconductors (Phys. Rev. B 110, 144516 (2024)).
3. **Difference between thin-flakes of $RbV_3Sb_5$ and $CsV_3Sb_5$**

   In our experiments, we have found very different behavior of $CsV_3Sb_5$ and $RbV_3Sb_5$ in the lowest temperature. Specifically speaking, at temperatures much lower than the critical temperature, the $CsV_3Sb_5$ shows no in-plane magnetic hysteresis as shown in Fig. S10, while $RbV_3Sb_5$ show clear unconventional magnetic hysteresis with critical magnetic field of forward sweep less than the backward sweep: $H_F < H_B$. Such observation further indicates that the superconducting pairing symmetry of $RbV_3Sb_5$ is probably distinct from $CsV_3Sb_5$. On the other hand, though hysteresis with conventional features of $H_F > H_B$ can be observed in the out-of-plane magnetic field dependence can be observed in both materials, the trivial origins should be carefully ruled out.

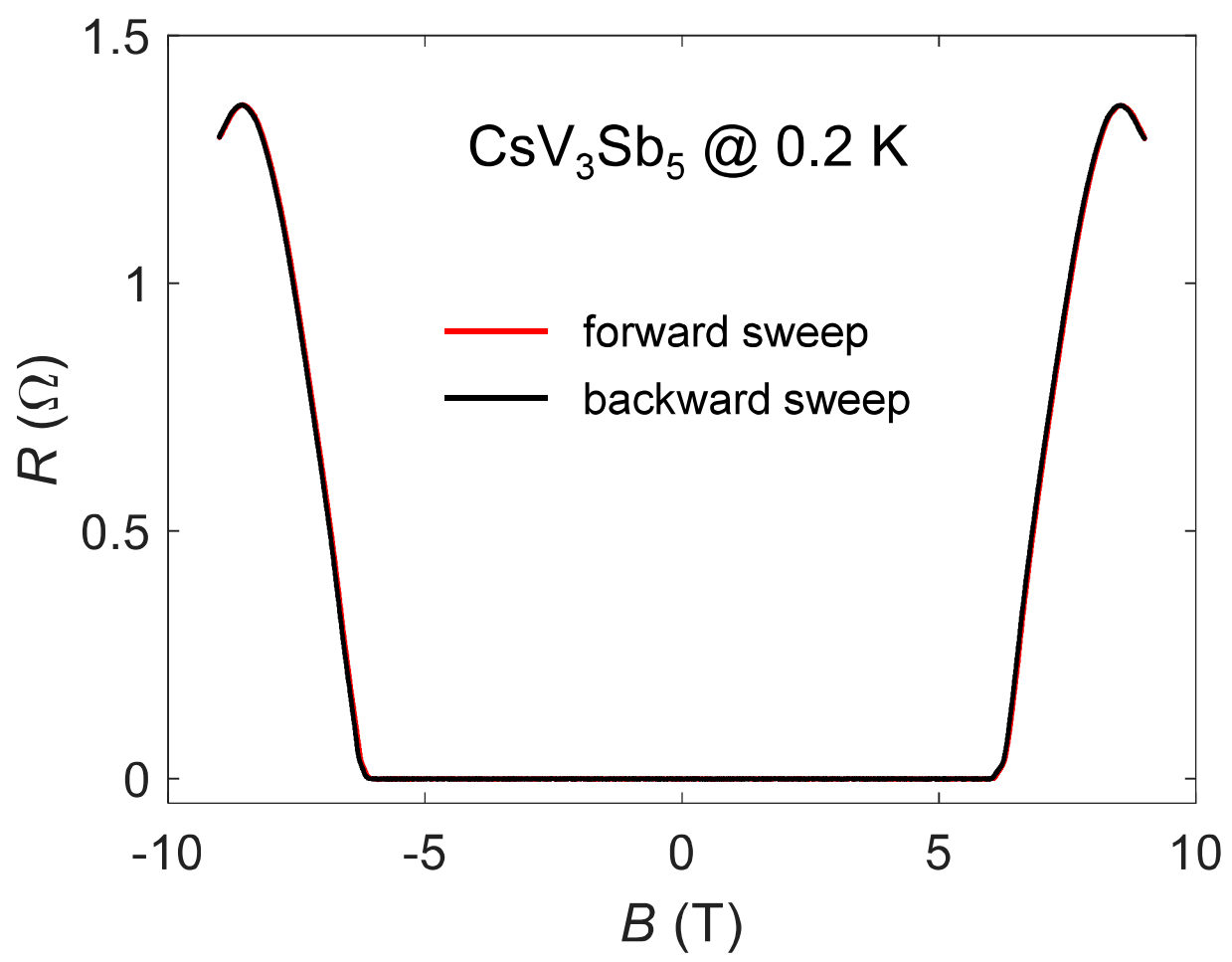

**Fig. S10 A typical magnetoresistance curve of thin-flake kagome material $CsV_3Sb_5$ below its superconducting transition temperature under an in-plane magnetic field sweep.** There is no hysteresis during the whole field-sweeping process.

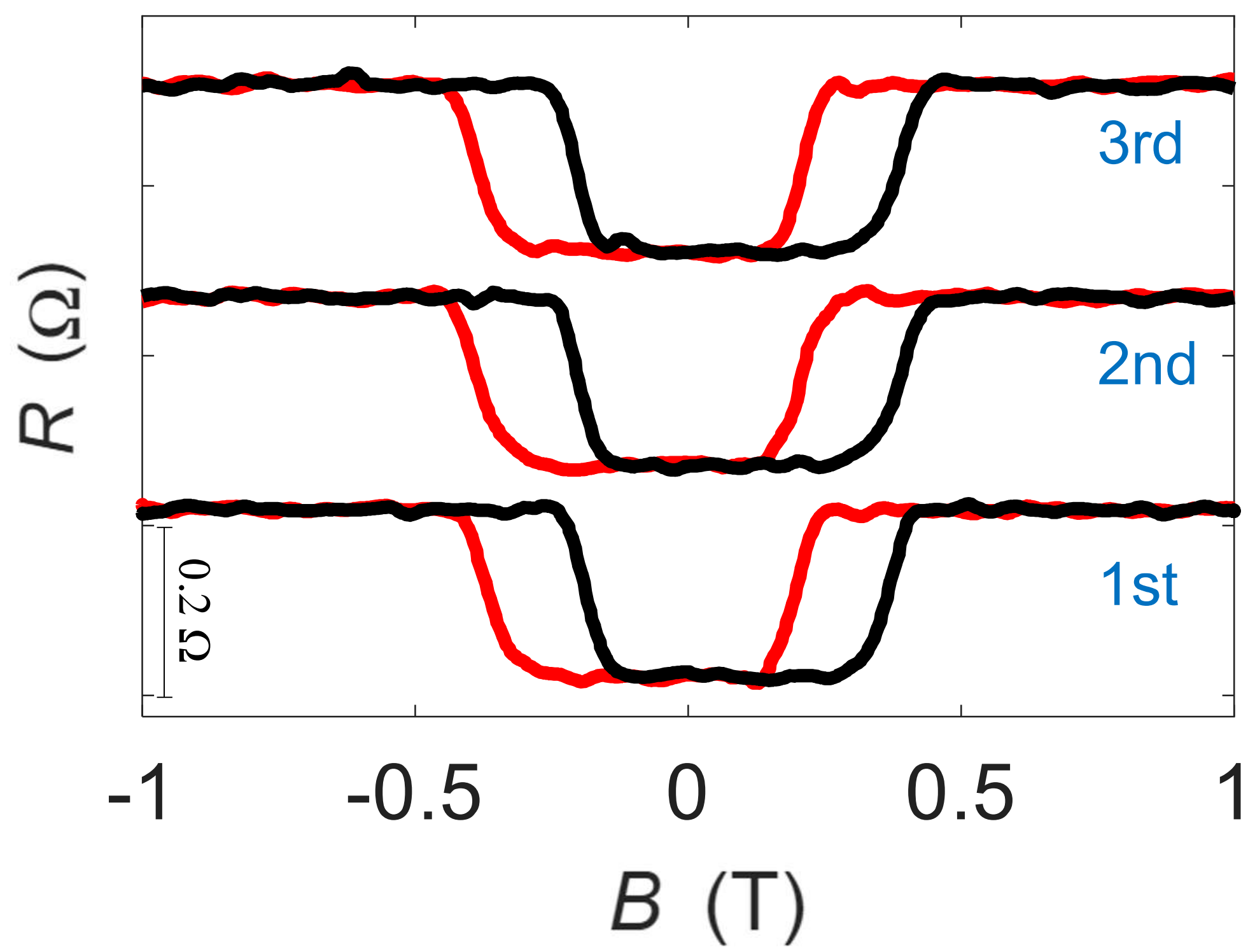

**Fig. S11.** The magnetoresistance curves of another sample show no memory effects since three consecutive magnetic field sweeps exhibit no discernible differences. The temperature is 150 mK.

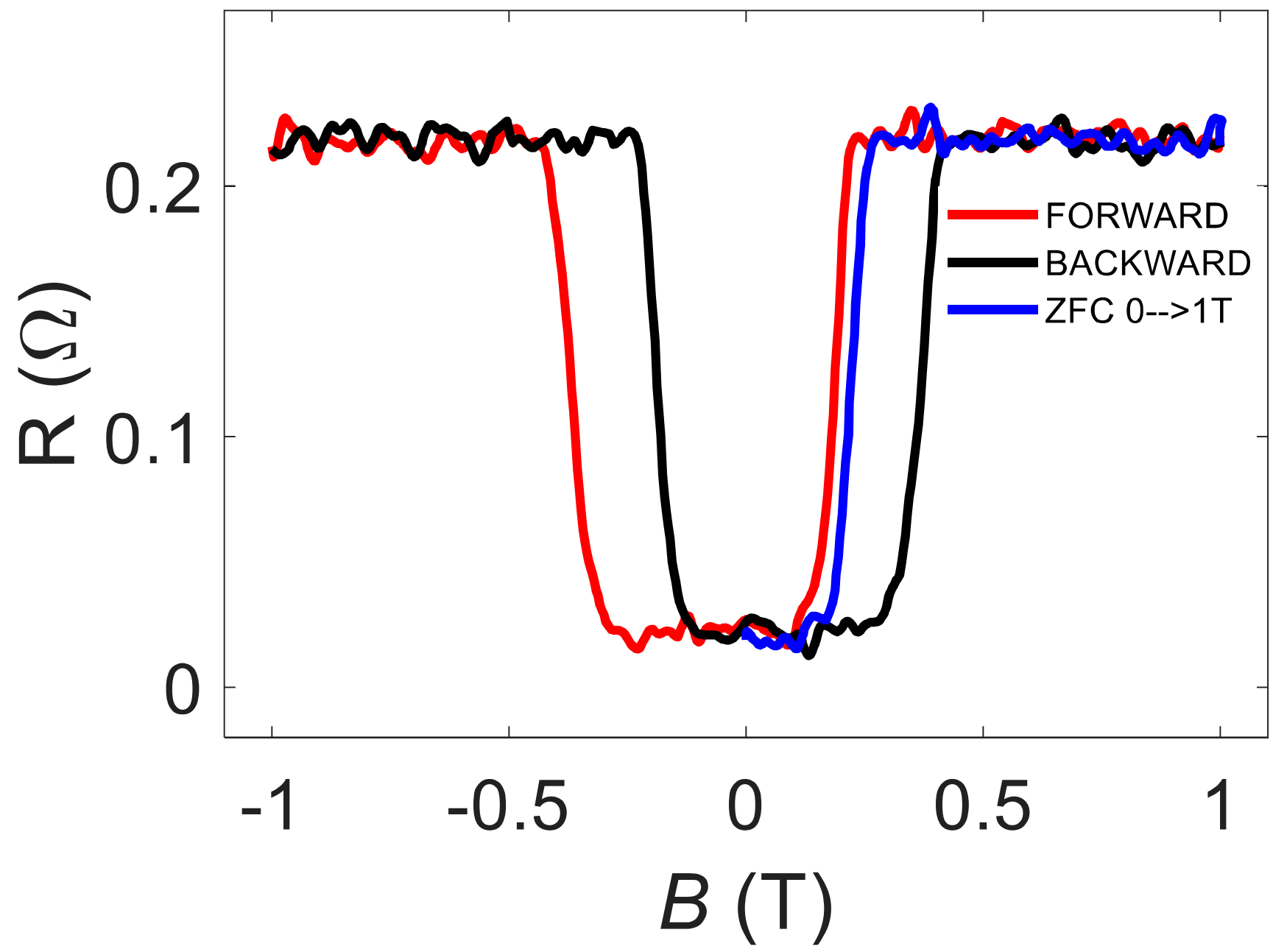

**Fig. S12** The magnetoresistance curve measured in field-sweeping process with three different initial magnetic fields: -1 T (red curve), 1 T (black curve) and 0 T (blue curve). The temperature is 150 mK.

# Section 2 The theory of spin-polarized p-wave superconductivity

## 2.1 The minimal four-orbital model of kagome superconductors

First of all, we construct the minimal model to describe the normal state of the kagome superconductor with local $d_{X^2-Y^2}$ orbitals belonging to the $V$ atoms and $p_z$ orbitals belonging to the $Sb$ atoms. The in-plane structure of the kagome superconductor is shown in Fig. S13. The Hamiltonian can be expressed as:

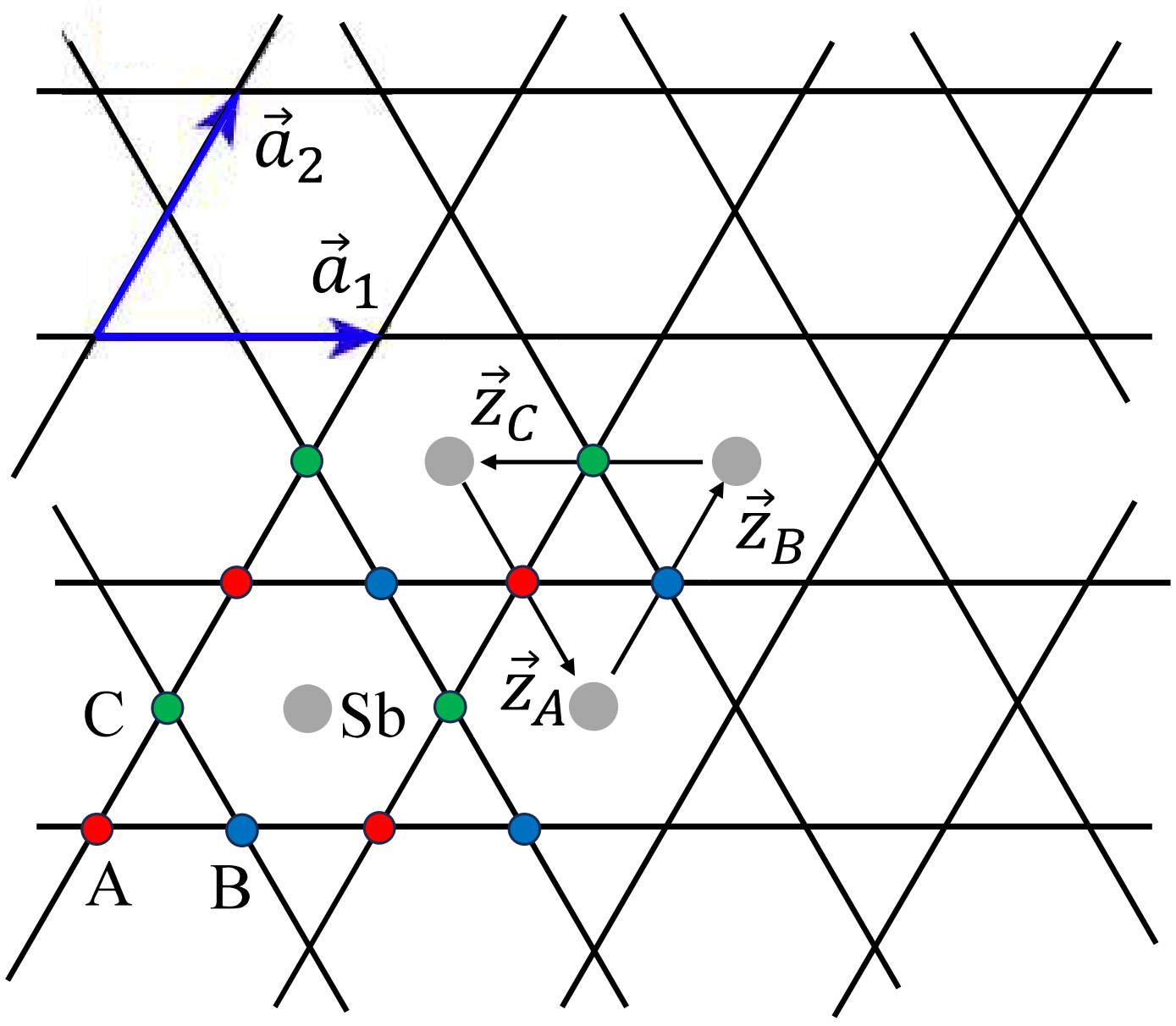

Fig. S13: The sketch illustrates the in-plane lattice structure of the Sb atoms (gray) and V atoms (R, B, G), with coordinates indicated by blue arrows in the figure. The z-axis of each local coordinate is depicted on each V atom.

$$H_{\vec{k}}^{4-orbital} = \begin{pmatrix} H_{\vec{k}\uparrow} & 0 & 0 & H_{\uparrow\downarrow} \\ 0 & H_{Sb\uparrow} & H_{\downarrow\uparrow}^{\dagger} & 0 \\ 0 & H_{\downarrow\uparrow} & H_{\vec{k}\downarrow} & 0 \\ H_{\downarrow\uparrow}^{\dagger} & 0 & 0 & H_{Sb\downarrow} \end{pmatrix}, \quad (1)$$

where the blocks are:

$$H_{\vec{k}\uparrow/\downarrow} = \begin{bmatrix} -\mu_V & -2t\cos(k_1/2) * nem & -2t\cos(k_2/2) \\ -2t\cos(k_1/2) * nem & -\mu_V & -2t\cos(k_3/2) \\ -2t\cos(k_2/2) & -2t\cos(k_3/2) & -\mu_V \end{bmatrix}, \quad (2)$$

$$H_{Sb\uparrow/\downarrow} = -2t_2(cos(k_1) + cos(k_2) + cos(k_3)) - \mu_{sb}, \tag{3}$$

$$H_{\uparrow\downarrow} = \begin{bmatrix} 2ibsin(\frac{k_2-k_1}{2})(-\frac{1}{2} + i\frac{\sqrt{3}}{2}) \\ 2ibsin(\frac{k_2}{2})(\frac{1}{2} + i\frac{\sqrt{3}}{2}) \\ 2ibsin(\frac{k_1}{2}) \end{bmatrix} \tag{4}$$

and

$$H_{\downarrow\uparrow} = \begin{bmatrix} 2ibsin(\frac{k_2-k_1}{2})(\frac{1}{2} + i\frac{\sqrt{3}}{2}) \\ 2ibsin(\frac{k_2}{2})(-\frac{1}{2} + i\frac{\sqrt{3}}{2}) \\ -2ibsin(\frac{k_1}{2}) \end{bmatrix}, \tag{5}$$

respectively. The basis of the Hamiltonian is:

$$\Psi_{\vec{k}} = (c_{\vec{k},V_A\uparrow}, c_{\vec{k},V_B\uparrow}, c_{\vec{k},V_C\uparrow}, c_{\vec{k},Sb,\uparrow}, c_{\vec{k},V_A\downarrow}, c_{\vec{k},V_B\downarrow}, c_{\vec{k},V_C\downarrow}, c_{\vec{k},Sb,\downarrow})^T. \tag{6}$$

The band structures, both without and with spin-orbital-parity coupling (SOPC) effect, are illustrated in Fig.1b of the main text. We conclude that the SOPC effect leads to the splitting of cross points in the original band structure of the kagome superconductors. The band near the Fermi surface will be used to construct an effective model to describe the physical properties of the kagome superconductor in the next section.

## 2.2 Effective single band model under magnetic field in manifestly covariant pseudospin basis (MCPB)

The pseudospin operators are constructed by projecting from orbital basis to band basis (the $\alpha_1$ and $\alpha_2$ represent the pseudospin up and down respectively):

$$\Phi_{\vec{k}} = (a_{\vec{k},n_1,\alpha_1}, a_{\vec{k},n_1,\alpha_2}, a_{\vec{k},n_2,\alpha_1}, a_{\vec{k},n_2,\alpha_2}, a_{\vec{k},n_3,\alpha_1}, a_{\vec{k},n_3,\alpha_2}, a_{\vec{k},n_4,\alpha_1}, a_{\vec{k},n_4,\alpha_2})^T. \tag{7}$$

The basis transformation matrix $U_{\vec{k}}$ satisfies the condition:

$$\begin{aligned} U_{\vec{k}} H^{4-orbital}_{\vec{k}} U^{\dagger}_{\vec{k}} &= H^{diag}_{\vec{k}} \\ U_{\vec{k}} \Psi_{\vec{k}} &= \Phi_{\vec{k}}. \end{aligned} \tag{8}$$

Since the properties of the superconductivity are dominated by the electrons near the Fermi surface, we project the orbitals to the second lowest band, the projector is $\tilde{U}_{\vec{k}}$. The target space of this projection is known as the manifestly covariant pseudospin basis (MCPB).

$$\tilde{U}_{\vec{k}} \Psi_{\vec{k}} = \tilde{\Phi}_{\vec{k}} = (a_{\vec{k},n_2,\alpha_1}, a_{\vec{k},n_2,\alpha_2})^T. \tag{9}$$

The $E_{n_2}(\vec{k})$ corresponds to the second lowest band, as depicted on the right side of Fig.1b in the main text. The projection can be expressed as:

$$\sigma'_i(\vec{k}) = \tilde{U}_{\vec{k}}(I_{4\times4} \bigotimes \sigma_i)\tilde{U}^{\dagger}_{\vec{k}}, \tag{10}$$

where $i = x, y, z$, $I_{4\times4}$ is identity matrix and $\sigma_i$ is Pauli matrix. The $\sigma'_i$ is the Pauli matrix in projecting space. Then, we can transform these matrices into $\sigma'_z(\vec{k})$-representation, the transformation matrix is $\Lambda_{\vec{k}}$.

$$\begin{aligned} \Lambda^{\dagger}_{\vec{k}} \sigma'_z(\vec{k}) \Lambda_{\vec{k}} &= \tilde{\sigma_z}(\vec{k}) = \lambda_{\vec{k}} \rho_z \\ \Lambda^{\dagger}_{\vec{k}} \sigma'_x(\vec{k}) \Lambda_{\vec{k}} &= \tilde{\sigma_x}(\vec{k}) = \sum_j a_{1j}(\vec{k}) \rho_j \\ \Lambda^{\dagger}_{\vec{k}} \sigma'_y(\vec{k}) \Lambda_{\vec{k}} &= \tilde{\sigma_y}(\vec{k}) = \sum_j a_{2j}(\vec{k}) \rho_j, \end{aligned} \tag{11}$$

where the $\rho_j(j = x, y, z)$ are Pauli matrices. So when the magnetic field $\vec{B} = (B_x, B_y, B_z)$ applied, the effective single band Hamiltonian can be written as:

$$H_0 = \sum_{\vec{k}\alpha} \xi_{\vec{k}} c^{\dagger}_{\vec{k}\alpha} c_{\vec{k}\alpha} + \mu_B \sum_{\vec{k},\alpha,\beta} c^{\dagger}_{\vec{k}\alpha} B_i a_{ij}(\vec{k}) \rho_j^{\alpha\beta} c_{\vec{k}\beta}. \tag{12}$$

Here we define an effective magnetic field: $g_j(\vec{k}) = \sum_i B_i a_{ij}(\vec{k})$ which acts on the pseudospin. For example, if $\vec{B}$ is $(0, B_y, 0)$, the effective field can be written as:

$$\vec{g}(\vec{k}) = (B_y a_{21}, B_y a_{22}, B_y a_{23}). \tag{13}$$

Then we define a new parameter:

$$\gamma_i(\vec{k}) = 2(a_{i1}^2 + a_{i2}^2 + a_{i3}^2). \tag{14}$$

as in Ref.[2]. Thus, the amplitude of the effective magnetic filed is $|\vec{g}(\vec{k})| = B_y\sqrt{\frac{\gamma_2}{2}}$.

## 2.3 The gap equation of the superconductivity

The Hamiltonian of the normal state in MCPB is:

$$H_0(\vec{k}) = \sum_{\vec{k}\alpha} \xi_{\vec{k}} c^{\dagger}_{\vec{k}\alpha} c_{\vec{k}\alpha} + \mu_B \sum_{\vec{k},\alpha,\beta} c^{\dagger}_{\vec{k}\alpha} \vec{g}(\vec{k}) \cdot \vec{\rho}_{\alpha\beta} c_{\vec{k}\beta}. \tag{15}$$

For BdG Hamiltonian, we choose the basis:

$$\psi_{\vec{k}} = (c_{\vec{k}\alpha}, c_{\vec{k}-\alpha}, c^{\dagger}_{-\vec{k}\alpha}, c^{\dagger}_{-\vec{k}-\alpha})^T. \tag{16}$$

The gap equation can be written as[1]:

$$\Delta_{ss'}(\vec{k}) = -k_B T \frac{1}{N_k} \sum_{n,\vec{k}'} \sum_{s_1,s_2,s_3,s_4} V_{\vec{k},\vec{k}';s,s',s_1s_2} G^0_{s_1s_3}(\vec{k}', i\omega_n) \Delta_{s_3,s_4}(\vec{k}') G^0_{s_4s_2}(-\vec{k}', -i\omega_n)^T, \tag{17}$$

which is equivalent to the linearized Eliashberg equation. The Green function $G^0_{ss'}(\vec{k}, i\omega_n)$ can be expressed as:

$$G^0(\vec{k}, i\omega_n) = G_+(\vec{k}, i\omega_n)\rho_0 + \hat{\vec{g}}_{\vec{k}} \cdot \vec{\rho} G_-(\vec{k}, i\omega_n), \tag{18}$$

where $G_{\pm}(\vec{k}, i\omega_n) = \frac{1}{2}(\frac{1}{i\omega_n - \epsilon_{k+}} \pm \frac{1}{i\omega_n - \epsilon_{k-}})$ and $\epsilon_{\vec{k}\pm} = \xi_{\vec{k}} \pm \mu_B|\vec{g}_{\vec{k}}|$. $\hat{\vec{g}}_{\vec{k}}$ is the unit vector in the direction of the effective magnetic field. Since the inversion symmetry is not broken, the pseudospin singlet and triplet are separated. To describe the properties of the superconductivity observed in experiments, we choose the pseudospin triplet channel below.

The order parameter of the pseudospin triplet pairing can be expressed as:

$$\Delta(\vec{k}) = (\vec{d}(\vec{k}) \cdot \vec{\rho}) i\rho_y. \tag{19}$$

We substitute the Eq.18 and Eq.19 into Eq.17. By using two equations:

$$\begin{aligned} (\vec{a} \cdot \vec{\sigma})(\vec{b} \cdot \vec{\sigma}) &= (\vec{a} \cdot \vec{b})\sigma_0 + i(\vec{a} \times \vec{b}) \cdot \vec{\sigma}, \\ i\sigma_2 \vec{\sigma}^T &= -\vec{\sigma}(i\sigma_2), \end{aligned} \tag{20}$$

and the decoupling of the interaction:

$$V_{\vec{k},\vec{k}';s,s',s_1,s_2} = \sum_a v_a (i\vec{d}_a(k) \cdot \vec{\rho}\rho_2)_{s,s'} (i\vec{d}_a(\vec{k}') \cdot \vec{\rho}\rho_2)^{\dagger}_{s_1s_2}, \tag{21}$$

we can simplify the gap equation, here the subscript "$a$" denotes different pairing channels belonging to different irreducible representations. Then the gap equation becomes ($v_a \equiv \frac{1}{2}v$, $\vec{d_a} \equiv \vec{d}$):

$$\frac{N_k}{v} = k_B T \sum_n \sum_{\vec{k}} |\vec{d}(\vec{k})|^2 [(G_+G_+ - G_-G_-) - 2(|\hat{g}_k \cdot \hat{\vec{d}}(\vec{k})|^2 - 1)G_-G_- + (\hat{\vec{g}}_{\vec{k}} \cdot (i\hat{\vec{d}}(\vec{k}) \times \hat{\vec{d}}^*(\vec{k})))(G_+G_- + G_-G_+)], \tag{22}$$

where the $\hat{\vec{p}}$ means the unit vector in the $\vec{p}$ 's direction and the Green function is written by shorthand:

$$G_\pm G_\pm = G_\pm(\vec{k}, i\omega_n)G_\pm(-\vec{k}, -i\omega_n). \tag{23}$$

## 2.4 The anisotropic behavior of the upper critical field in single-layer limit

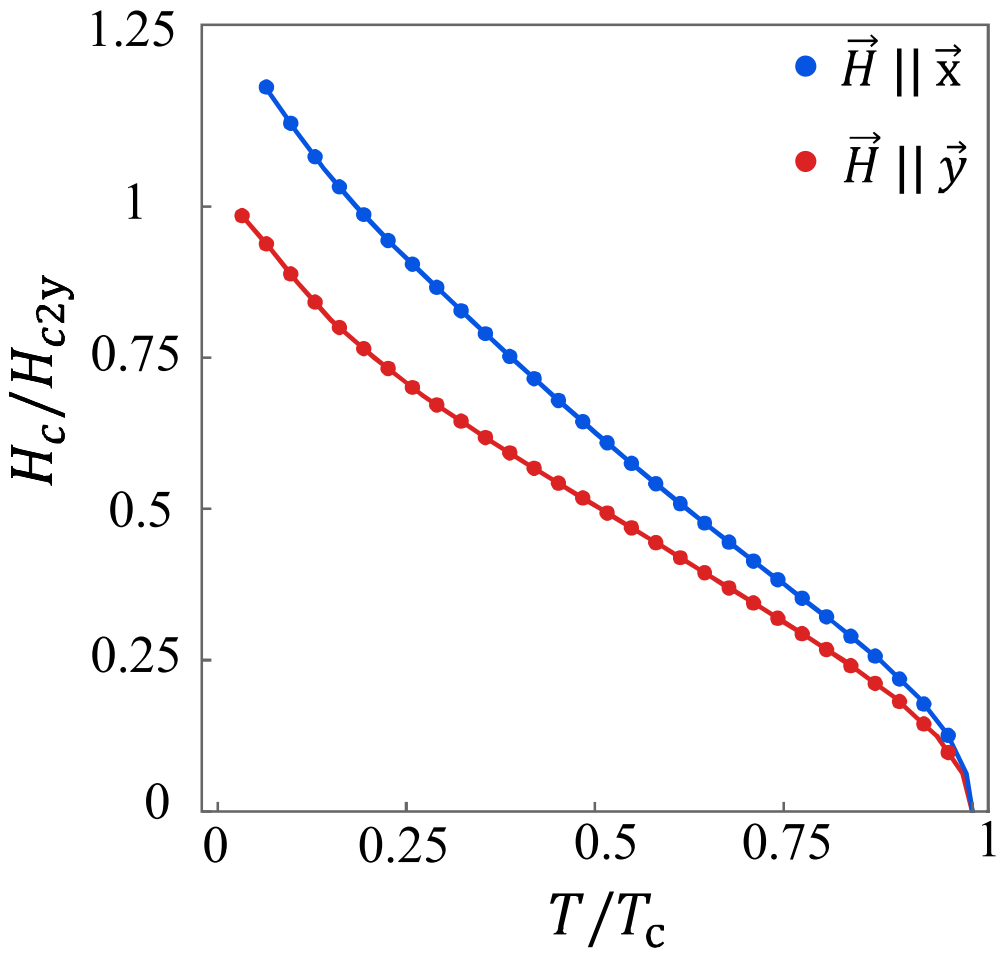

Fig. S14: The temperature-dependent upper critical field. The blue and red lines represent cases where the magnetic field is applied along the $x$-axis and $y$-axis, respectively

When the magnetic field is applied, the most favorable pseudospin triplet pairing state is determined by the equation: $i\hat{\vec{g}}_{\vec{k}} \cdot (\hat{\vec{d}}(\vec{k}) \times \hat{\vec{d}}^*(\vec{k})) > 0$. We assume that the decoupling coefficient in gap equation is a function of polarization parameter $\eta$, namely, $v = v_0(1 - (\eta - \eta_0)^2)^{-1}$, where the $\eta$ is $|cos(\xi)^2 - sin(\xi)^2|$ for $\vec{d} = sin(k_x)(cos(\chi)cos(\xi), sin(\chi)cos(\xi), isin(\xi))$. When $\eta = 1$, there is no spin polarization in the superconducting state, in other words, the pairing state is unitary. Conversely, when $\eta = 0$, half of the electrons remain unpaired, representing a fully polarized superconducting state. Therefore, the polarization of the superconducting state can be defined as $1 - \eta$. With this assumption, we can obtain a spontaneously polarized superconducting state, which is consistent with experimental observations. The parameter $\chi$ is determined by the coupling of the spin polarization with the effective magnetic field: $\hat{\vec{g}}_k \cdot \hat{\vec{d}}(\vec{k})$.

The upper critical field exhibit anisotropic behavior in single-layer limit when considering both the spin-orbital-parity-coupling effect and nematic normal state, as shown in Fig. S14 by the blue and red lines, which is also depicted in Fig. S3.

## 2.5 Open boundary calculation for nodal p-wave superconductivity

We choose the superconducting states with the order parameter belonging to the $B_{2u}$ and $B_{3u}$ irreducible representations to perform the open boundary calculation, respectively. Our aim is to demonstrate the Majorana zero modes located at the boundary of the superconductor, and the results are shown in Fig. S15.

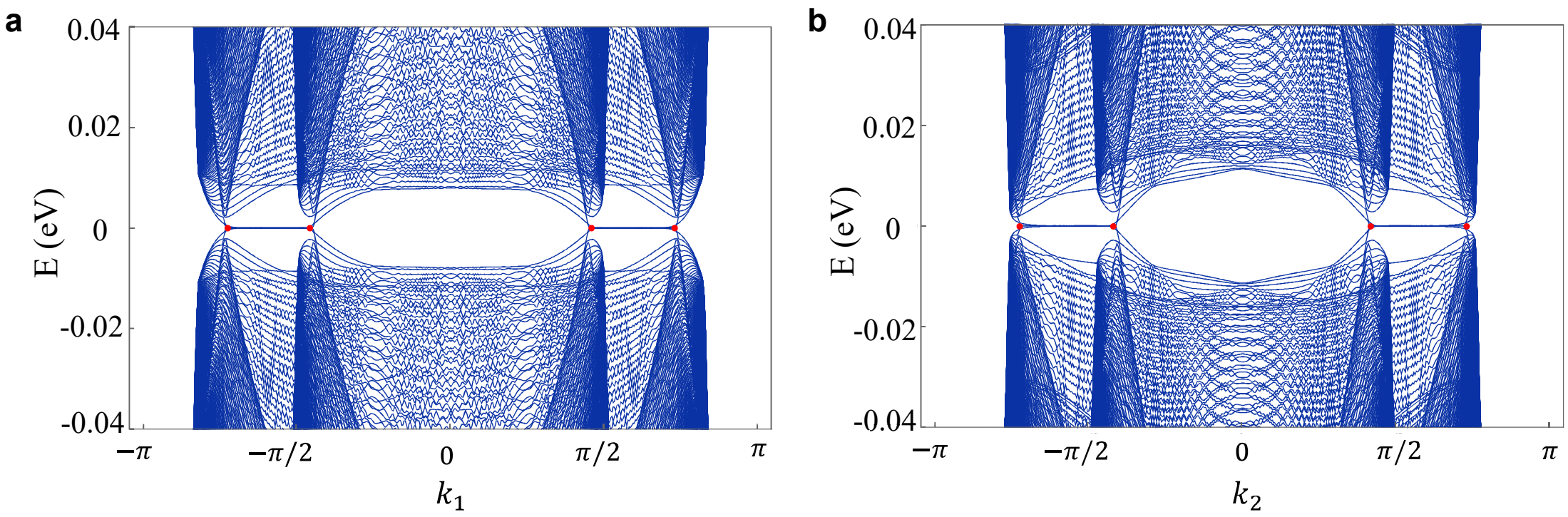

Fig. S15: **a**&**b**,The open boundary spectrum of the superconductivity with the order parameter belonging to the $B_{2u}$ and $B_{3u}$ irreducible representations, respectively. For these two cases, the open boundary directions are along the $\vec{a}_2$-axis and $\vec{a}_1$-axis, respectively. There are Majorana zero modes located between two point nodes belonging to different Fermi surface in both cases. The amplitude of the gap is chosen to be $\Delta_0 = 0.05$.

## 2.6 Parameters in supplementary material

In the 4-orbital model, we choose the parameters: $t = 1$, $t_2 = 0.51$, $\mu_V = -0.275$, $\mu_{Sb} = -2.3$, $b = 0.5$, $nem = 2$ (when considering the nematic effects). Then, we fit the one-band model on a triangular lattice to obtain the hopping parameters: $t_0 = 0.0656$, $tt = -0.2935$, $ttt = -0.0813$, and $\mu_{fit} = -0.3184$. The one-band model is:

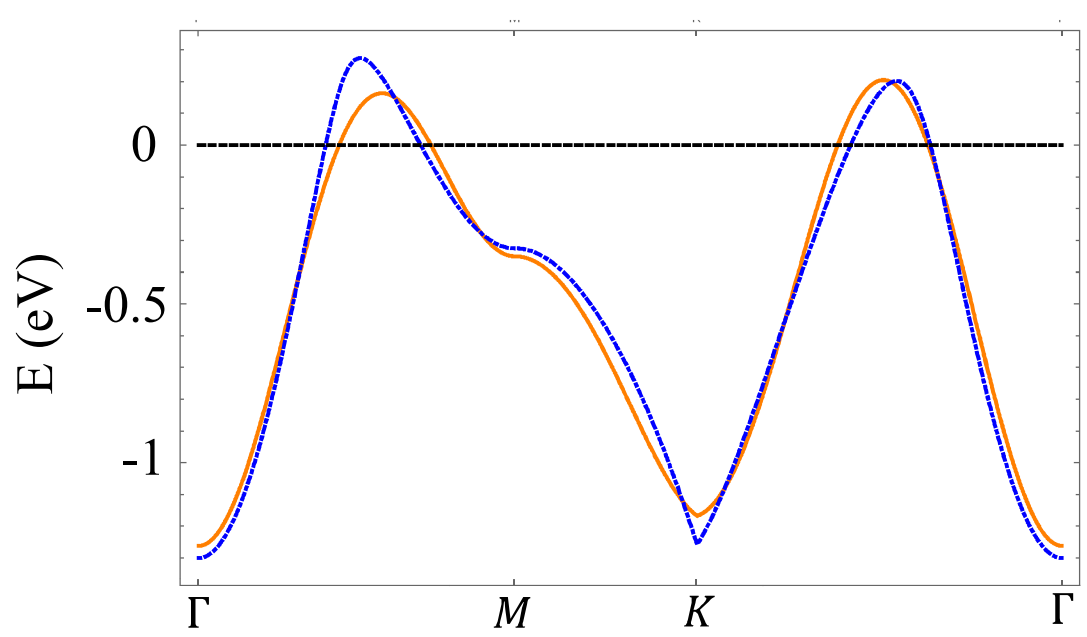

Fig. S16: The band structure of the effective one-band model, projected from the 4-orbital model, is shown as the dashed blue line. The orange line represents the model on a triangular lattice, with parameters determined through the fitting approach.

$$H_0(\vec{k}) = t_0(nem * cos(k_x) + cos(\frac{k_x}{2} + \frac{\sqrt{3}k_y}{2}) + cos(-\frac{k_x}{2} + \frac{\sqrt{3}k_y}{2})) + tt(cos(\sqrt{3}k_y + cos(\frac{3k_x}{2} + \frac{\sqrt{3}k_y}{2}) + cos(\frac{3k_x}{2} - \frac{\sqrt{3}k_y}{2})) + ttt(nem * cos(2k_x) + cos(k_x + \sqrt{3}k_y) + cos(-k_x + \sqrt{3}k_y)) + \mu_{fit}. \tag{24}$$

When we apply a magnetic field and solve the gap equation, the parameters are chosen as: $\eta_0 = |cos(\xi_0)^2 - sin(\xi_0)^2|$, where $\xi_0 = 0.3\pi$. Additionally, we choose the decoupling coefficient $\frac{1}{v_0} = 0.7028$ and the number of k points $N_k = 10000$.